\documentclass[twocolumn]{aastex631}

\newcommand{\rjup}{\mbox{$\mathcal{R}_\mathrm{J}^\mathrm{N}$}}
\newcommand{\mjup}{\mbox{$M_{\rm{Jup}}$}}
\newcommand{\teff}{\mbox{$T_{\mathrm{eff}}$}} 
\newcommand{\logg}{\mbox{log $g$}}
\newcommand{\fbol}{$F_\mathrm{bol}$}
\newcommand{\lbol}{$L_\mathrm{bol}$}

\newcommand{\kzz}{\mbox{$K_{\mathrm{zz}}$}} 

\newcommand{\mic}{\mbox{$\mu$m}} 
\newcommand{\wname}{WISE~0359$-$54}
\newcommand{\wfname}{WISE~J035934.06$-$540154.6}
\graphicspath{{./}{figures/}}

\begin{document}

\title{The First \textit{JWST} Spectral Energy Distribution of a Y dwarf}
\shorttitle{First \textit{JWST} Spectral Energy Distribution of a Y dwarf}
\author[0000-0002-6721-1844]{Samuel A. Beiler}
\affiliation{Ritter Astrophysical Research Center, Department of Physics \& Astronomy,
University of Toledo, 2801 W. Bancroft St.,
Toledo, OH 43606, USA}

\author[0000-0001-7780-3352]{Michael C. Cushing}
\affiliation{Ritter Astrophysical Research Center, Department of Physics \& Astronomy,
University of Toledo, 2801 W. Bancroft St.,
Toledo, OH 43606, USA}

\author[0000-0002-6721-1844]{J. Davy Kirkpatrick}
\affiliation{IPAC, Mail Code 100-22, Caltech, 1200 E. California Boulevard, Pasadena, CA 91125, USA}

\author[0000-0003-4269-260X]{Adam C. Schneider}
\affiliation{United States Naval Observatory, Flagstaff Station, 10391 West Naval Observatory Road, Flagstaff, AZ 86005, USA}

\author[0000-0003-1622-1302]{Sagnick Mukherjee}
\affiliation{Department of Astronomy and Astrophysics, University of California, Santa Cruz, 1156 High Street, Santa Cruz, CA 95064, USA}

\author[0000-0002-5251-2943]{Mark S. Marley}
\affiliation{Lunar and Planetary Laboratory, University of Arizona, 1629 E. University Boulevard, Tucson, AZ 85721, USA}

\begin{abstract}
We present the first \textit{JWST} spectral energy distribution of a Y dwarf. This spectral energy distribution of the Y0 dwarf \wfname{} consists of low-resolution ($\lambda$/$\Delta\lambda$ $\sim$ 100) spectroscopy from 1--12 \mic{} and three photometric points at 15, 18, and 21 \mic. The spectrum exhibits numerous fundamental, overtone, and combination rotational-vibrational bands of H$_2$O, CH$_4$, CO, CO$_2$, and NH$_3$, including the previously unidentified $\nu_3$ band of NH$_3$ at 3 \mic. Using a Rayleigh-Jeans tail to account for the flux emerging at wavelengths greater than 21 \mic, we measure a bolometric luminosity of $1.523\pm0.090\times10^{20}$ W. We determine a semi-empirical effective temperature estimate of $467^{+16}_{-18}$ K using the bolometric luminosity and evolutionary models to estimate a radius. Finally, we compare the spectrum and photometry to a grid of atmospheric models and find reasonably good agreement with a model having \teff=450 K, \logg=3.25 [cm s$^{-2}$], [M/H]=$-0.3$. However, the low surface gravity implies an extremely low mass of 1 \mjup{} and a very young age of 20 Myr, the latter of which is inconsistent with simulations of volume-limited samples of cool brown dwarfs. 
\end{abstract}


\section{Introduction} \label{sec:intro}
Y dwarfs are the coolest products of stellar formation, with effective temperatures less than 600 K \citep{kirkpatrick_field_2021} and spectral energy distributions that peak at $\sim$5 \mic{} in $f_\lambda$ space. Within the cool atmospheres of Y dwarfs are an abundance of molecular species, creating deep absorption features and forcing more light to escape in the mid-infrared atmospheric windows. The hotter L and T dwarfs were studied at these wavelengths spectroscopically using data from the \textit{Spitzer Space Telescope} \citep{werner_spitzer_2004} and \textit{AKARI} \citep{murakami_infrared_2007}, but \textit{AKARI} was not sensitive enough to study Y dwarfs and the \textit{Spitzer} cryogenic mission ended before the discovery of Y dwarfs \citep{cushing_discovery_2011}. As such, Y dwarf observations have been mostly limited to spectra shortward of 1.7 \mic{} and a couple of \textit{Spitzer} and \textit{Wide-field Infrared Survey Explorer} \citep[\textit{WISE,}][]{wright_wide-field_2010} photometric measurements at longer wavelengths \citep[e.g.][]{schneider_hubble_2015,leggett_y-type_2017}. There are some ground-based spectra from 2--5 \mic{} \citep[e.g.][]{skemer_first_2016,morley_l_2018, miles_observations_2020}, but even with these near-heroic observational efforts the signal-to-noise and resolution is limited by both telluric absorption and emission.

The launch of \textit{JWST} \citep{rigby_science_2023} opens up the entire wavelength interval, from the far optical through the mid-infrared (0.6--24 \mic), where the bulk of Y dwarf emission is predicted to be observed. Using \textit{JWST} we are building the spectral energy distributions of 24 late T and Y dwarfs with well-measured parallaxes that consist of low resolution spectra from $\sim$1--12 \mic{} and photometry out to 21 \mic{}. This wavelength range contains absorption bands from all of the dominant carbon, nitrogen, and oxygen bearing species found in the atmospheres of cool brown dwarfs including $\mathrm{H}_2\mathrm{O}$, $\mathrm{CH}_4$, $\mathrm{NH}_3$, $\mathrm{CO}$, and $\mathrm{CO}_2$. In addition, direct integration of the spectral energy distributions yield \fbol, which can then be used to compute the bolometric luminosity (with a known distance) and effective temperature (with a suitable radius).  

In this paper, we present the first \textit{JWST} $\sim$1--21 \mic{} spectral energy distribution of a Y dwarf, \wfname{} (hereafter \wname). In $\S$\ref{sec:obvs} and $\S$\ref{sec:reduc} we discuss observations and reduction of the raw data. In $\S$\ref{sec:spec} we present the spectrum, and discuss the major molecular features. In $\S$\ref{sec:prop} we construct a flux calibrated spectral energy distribution and calculate the bolometric flux, bolometric luminosity, and effective temperature of \wname. In $\S$\ref{sec:fits} we fit models to our spectrum to compare our derived effective temperature with that of the best-fit model spectrum.

\section{Observations} \label{sec:obvs}
We observed \wname{} using \textit{JWST}'s Near Infrared Spectrograph \citep[NIRSpec,][]{jakobsen_near-infrared_2022} and Mid Infrared Instrument \citep[MIRI, ][]{rieke_mid-infrared_2015}, collecting low-resolution spectroscopy with both instruments, and broad-band photometry with MIRI (GO 2302, PI Cushing). Details of our observations can be found in Table \ref{tbl:obs}. The observations were carried out sequentially across 2.5 hours on UT 2022-Sep-12 in the following order: NIRSpec, MIRI spectroscopy, MIRI photometry. 

NIRSpec was used in fixed-slit mode with the CLEAR/PRISM filter and the S200A1 slit ($0\farcs2\times3\farcs2$). This filter allowed us to obtain a spectrum from 0.6--5.3 \mic{} with a spectral resolving power ($R = {\lambda}/{\Delta \lambda}$) of $\sim$100. The observations were completed with the NRSRAPID readout pattern and a 5 point dither, for a total exposure time of 85.792 s. 

MIRI spectroscopy was carried out using the low-resolution spectrometer (LRS) in slit mode ($0\farcs51\times4\farcs7$). MIRI LRS allowed us to obtain a spectrum from 5--12 \mic{} with a resolving power of $\sim$100. We used the FASTR1 readout mode and a 2-point ``along slit nod" dither, and observed for a total exposure time of 2775.04~s. Both the MIRI and NIRSpec slit land on an area of the sky with no background \textit{WISE} W1 (3.4 \mic) or W2 (4.6 \mic) detections.

We also collected photometry in four MIRI broad-band filters: F1000W, F1500W, F1800W, and F2100W ($\lambda_\mathrm{pivot}$= 9.954, 15.065, 17.987, 20.795 \mic{} respectively). For the observations we used the FASTR1 readout, and a 4-point dither, with total exposure times of (in increasing wavelength) 77.701, 77.701, 199.803, and 954.614 s. The order of the observations was F1500W, F1800W, F2100W, and F1000W.

\begin{deluxetable*}{lccccrcr}
\tablecaption{\textit{JWST} Observations of \wfname \label{tbl:obs}}
\tablehead{
\colhead{Obs. Start Time (UT)} &
\colhead{Instrument} &
\colhead{Mode} &
\colhead{Subarray} &
\colhead{Readout} &
\colhead{Wavelength} &
\colhead{$\lambda/\Delta\lambda$} &
\colhead{N $\times$ Int. Time (s)}
}
\startdata
2022-09-12 21:12:04& NIRSpec& CLEAR/PRISM & SUBS200A1 & NRSRAPID & 0.6--5.3 \mic & 30--300 & $5\times 17.158$ \\
2022-09-12 21:47:19& MIRI& LRS& FULL & FASTR1& 5.0--12.0 \mic& 50--200 & $2\times1387.530$\\
2022-09-12 22:52:01& MIRI& F1500W& FULL & FASTR1& 15.064 \mic& -- & $4\times19.425$\\
2022-09-12 23:00:43& MIRI& F1800W& FULL & FASTR1& 17.894 \mic& -- & $4\times49.951$\\
2022-09-12 23:12:09& MIRI& F2100W& FULL & FASTR1& 20.795 \mic& -- & $12\times79.551$\\
2022-09-12 23:36:14& MIRI& F1000W& FULL & FASTR1& 9.909 \mic & -- & $4\times19.425$ \\
\enddata
\end{deluxetable*}

\section{Data Reduction} \label{sec:reduc}
\subsection{\textit{JWST} Pipeline and Wavelength Calibration}
We use the official \textit{JWST} pipeline (Version 1.8.2) for the data reduction, using the 11.16.14 CRDS (Calibration Reference Data System) version and 1027.pmap CRDS context to assign the reference files. The pipeline is split into three stages. Stage 1 takes the detector ramps of an exposure, makes various detector-level corrections (e.g. bias subtraction, dark subtraction, cosmic ray detection), and then fits the ramp slopes to generate a count-rate image. Stage 2 takes these individual count-rate images and assigns the world-coordinate system, performs instrument-level and observing mode corrections, and performs absolute flux calibration on the images. These calibrations are derived from a combination of pre-flight tests and in-flight observations. Stage 3 then takes all the exposures for a single observation and aligns them, flags outlier pixels via sigma clipping, and combines the exposures into a single image. Stage 3 outputs either an extracted 1D spectrum for spectroscopic observations, or a source catalog of fluxes and magnitudes for imaging observations. The photometric fluxes and magnitudes are in Table \ref{tbl:prop}. 

There are three modifications to the standard reduction steps for our object:

1. The ``extract\_1D'' step of the Stage 3 pipeline assumes the target will be at its nominal position, however the actual position is offset from this location, most likely a result of the uncertainty in the parallax ($73.2 \pm 2.0$ mas) and proper motion ($\mu_\mathrm{R.A.}=-134.1 \pm 0.7$ mas yr$^{-1}$ and $\mu_\mathrm{Decl.}=-758.9 \pm 0.9$ mas yr$^{-1}$, \citealt{kirkpatrick_field_2021}) of the source. As such we require the ``extract\_1D'' step to extract from the center of the point spread function with the default extraction radius. 

2. The uncertainty in the zero point is not included in the measured photometry as the absolute calibration for MIRI photometry in the current version of the pipeline is incomplete \citep{gordon_james_2022}. We therefore, based off the instrument requirements, choose to assume a 5\% uncertainty for all magnitudes regardless of the signal-to-noise of the observations. 

3. There is a known issue with the MIRI LRS wavelength calibration, causing a $\sim$0.02--0.05 \mic{} inaccuracy in the reported wavelength values\footnote{Details at: https://jwst-docs.stsci.edu/jwst-calibration-pipeline-caveats/jwst-miri-lrs-pipeline-caveats}. This creates a noticeable offset when comparing our spectrum to models and molecular opacities. 

\begin{figure}
\includegraphics[width=.47\textwidth]{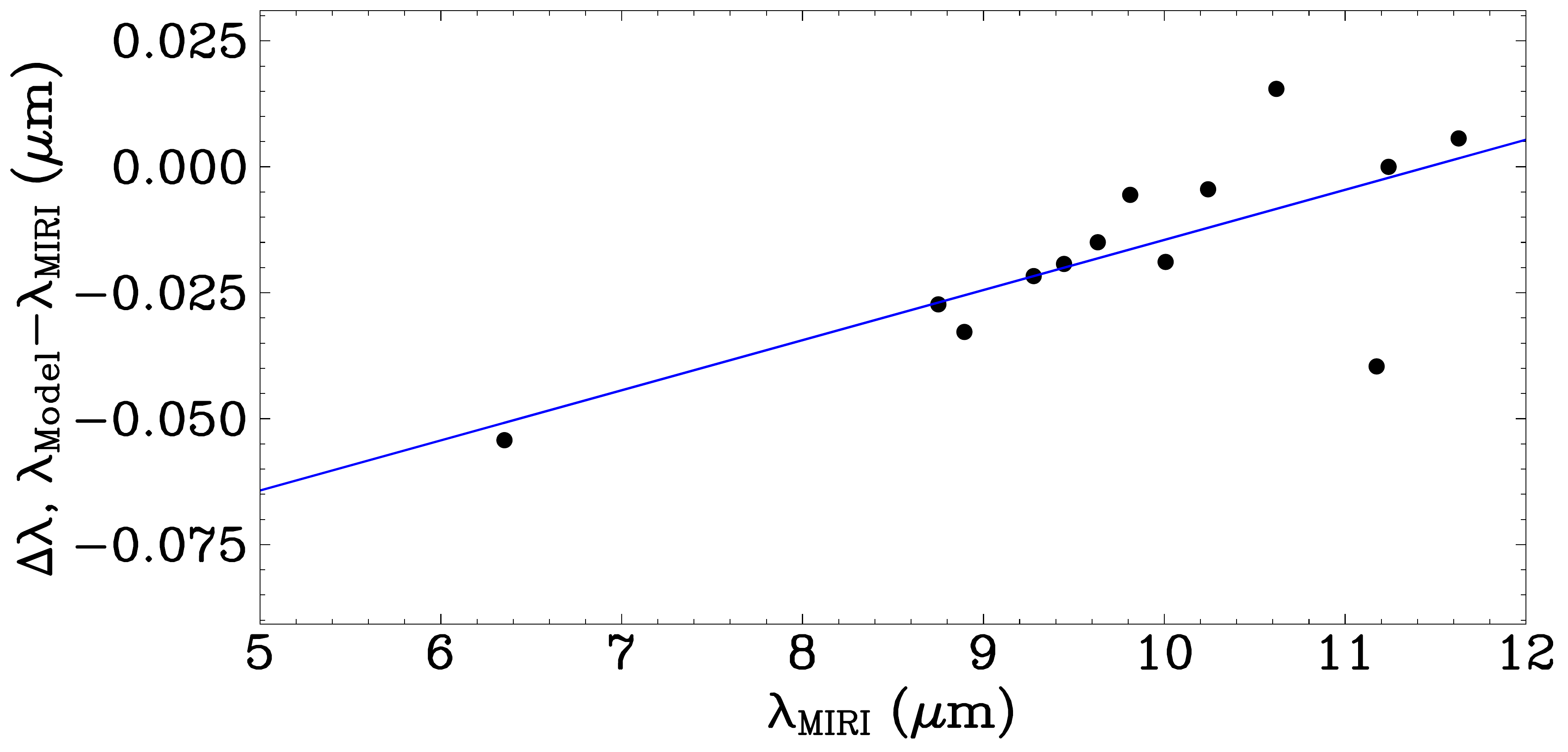}
\centering
\caption{The offset of the model spectrum's fitted peaks and observed spectrum's peaks as a function of the observed wavelength (black), with a linear fit that we used to correct the MIRI wavelength calibration.}\label{fig:wavecal}
\end{figure}

To correct this we measure the wavelength offset between our data and a 500 K model spectrum \citep[Sonora Cholla;][]{karalidi_sonora_2021} by comparing the central wavelengths of Gaussian fits to 13 features. Specifically, we fit to the peak in the middle of the water feature at 6.35 \mic, and the peaks of the ammonia features from 8.6--11.7 \mic{} (excluding those from 8.96--9.2 \mic). The peaks that were not included are either poorly approximated by a Gaussian or have too few data points in the spectrum to constrain the Gaussian fit. The best linear fit to the offset between the model and data central wavelengths ($\Delta\lambda$, $\lambda_\mathrm{Model}-\lambda_\mathrm{MIRI}$ in \mic{}) as a function of the observed wavelength is $\Delta\lambda=0.0106\lambda_{\mathrm{MIRI}} - 0.120$, which can be applied to correct the wavelength of each data point. This fit is shown in Figure \ref{fig:wavecal}. While there are few peaks to be fit at the blue end of the MIRI LRS spectrum, the offset seems well corrected by eye at wavelengths between 5--9 \mic{} with this fit.

\begin{deluxetable}{lcr}
\tablecaption{Astrometric and Photometric Properties of \wname \label{tbl:prop}}
\tablehead{
\colhead{Property} &
\colhead{Value} &
\colhead{Reference}
}
\startdata
Spectral Type & Y0 & 2\\
Parallax  $ \varpi $ (mas)& $73.2\pm2.0$ &3\\
Distance (pc)  & $13.57\pm0.37$ & 3\\
$\mu_\mathrm{R.A.}$ & $-134.1 \pm 0.7$ mas yr$^{-1}$& 3\\
$\mu_\mathrm{Decl.}$ & $-758.9 \pm 0.9$ mas yr$^{-1}$ & 3\\
IRAC [3.6] (mag) & $17.55\pm0.072$& 2\\
IRAC [4.5] (mag) & $15.32\pm0.023$& 2\\
MIRI F1000W (mJy)\tablenotemark{a} & $0.1342\pm0.0067$ &  1\\ 
MIRI F1500W (mJy)\tablenotemark{a} & $0.1231\pm0.0062$& 1\\ 
MIRI F1800W (mJy)\tablenotemark{a} & $0.1019\pm0.0051$ & 1\\ 
MIRI F2100W (mJy)\tablenotemark{a} & $0.0731\pm0.0037$ & 1\\ 
MIRI F1000W (mag)\tablenotemark{a} & $13.62\pm0.054$ & 1\\ 
MIRI F1500W (mag)\tablenotemark{a} & $12.83\pm0.054$ & 1\\ 
MIRI F1800W (mag)\tablenotemark{a} & $12.65\pm0.054$ & 1\\ 
MIRI F2100W (mag)\tablenotemark{a} & $12.71\pm0.054$ & 1\\ 
\enddata
\tablenotetext{a}{All MIRI photometric points are assumed to have a 5\% uncertainty in flux density.}
\tablerefs{
(1)~This paper, (2)~\citet{kirkpatrick_further_2012},
(3)~\citet{kirkpatrick_field_2021}.}
\end{deluxetable}

\subsection{Flux Calibration and Merging of Spectra} \label{subsec:reducCali}
The pre-flight goal for the precision of the absolute flux calibration of \textit{JWST} spectra was $\sim10$\% \citep{gordon_james_2022}. To obtain a relative precision of 3\% on the bolometric flux measurement, we obtained photometry to improve the absolute calibration to $\sim$5\%. The NIRSpec spectrum was calibrated with \textit{Spitzer/IRAC} Channel 2 ([4.5], 4.5 \mic) photometry \citep{kirkpatrick_further_2012}, which was shown to have no significant variability for \wname{} by \citet{brooks_long-term_2023}. The MIRI LRS spectrum was calibrated with MIRI F1000W (10 $\mu$m) photometry, which were observed within an hour of each other. Both of these photometric values and uncertainties can be found in Table {\ref{tbl:prop}. From these flux densities we calculate the scaling factors needed to convert our spectra and their errors to absolute units of Jansky \citep{reach_absolute_2005,gordon_james_2022}. It is worth noting \textit{Spitzer} and \textit{JWST} photometry assume different nominal spectra, $\nu f_{\nu}$ = constant and $f_\lambda$ = constant respectively. 

We create a continuous 1--12 \mic{} spectrum by merging the NIRSpec and MIRI spectra at their overlap from 5 to 5.3 \mic. The NIRSpec spectrum has a higher resolving power than the MIRI LRS spectrum at these wavelengths, allowing us to resolve several absorption features that are not seen with MIRI. To balance the trade-off between including these spectral features and including low signal-to-noise data, we cut the NIRSpec spectrum where it first drops below a signal-to-noise of 10 at 5.14 \mic. Similarly, due to the low flux shortward of the $Y$ band, we made a cut at the last point before the $Y$ band with a negative flux ($<$0.96 \mic). Our final reduced spectrum and photometry are presented in Figure \ref{fig:speclabel}.

\begin{figure*}
\includegraphics[width=\textwidth]{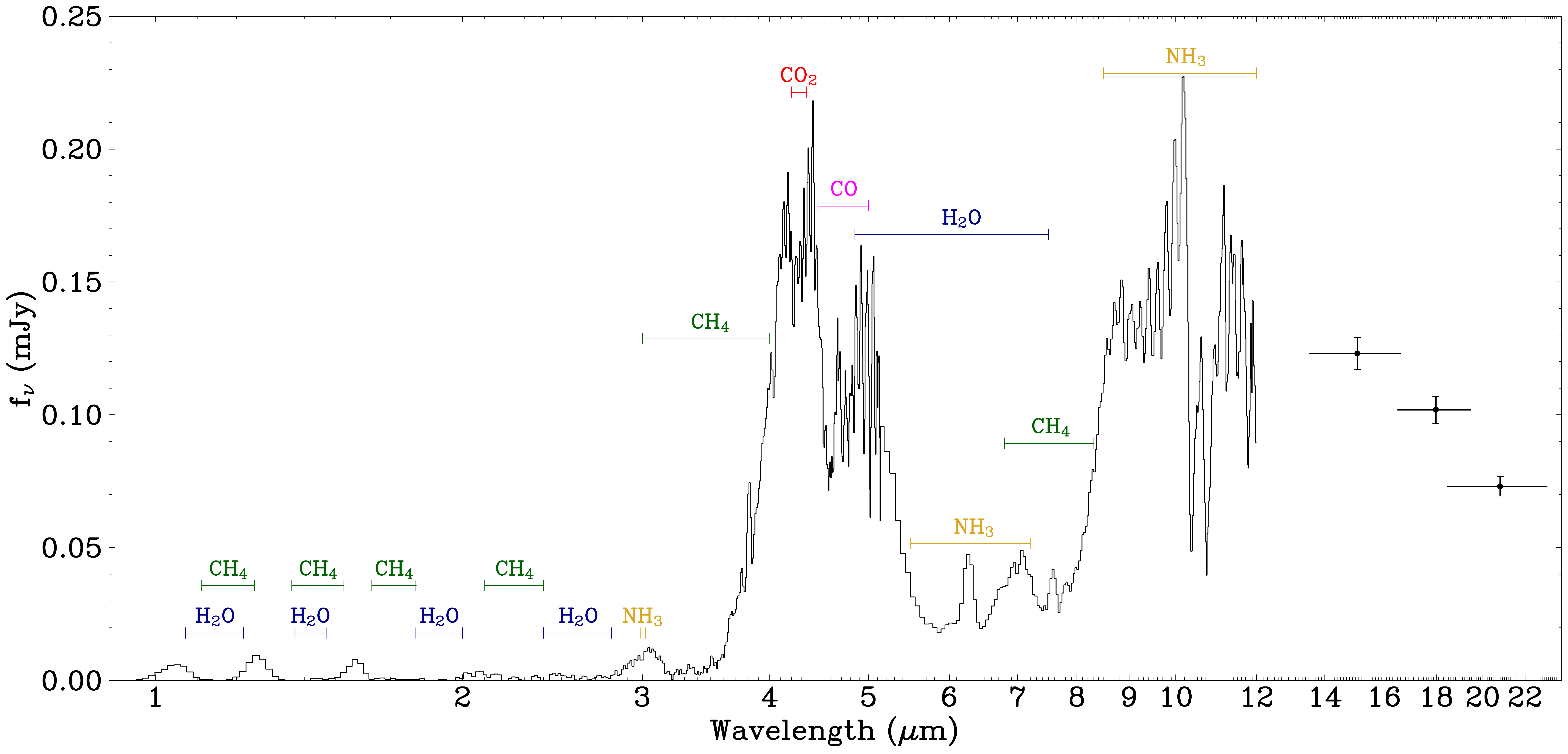}
\centering
\caption{The first full spectral energy distribution (black) from 1--21 \mic{} of a Y dwarf (\wname), with key molecular features identified (color). It consists of merged and calibrated \textit{JWST} spectra and photometry (F1500W, F1800W, and F2100W), in dimensions of $f_\nu$. The transition between the NIRSpec and MIRI LRS spectra occurs at 5.14 \mic{} and can be seen easily by the change in resolving power. The FWHM of the photometric bandpasses are marked with horizontal bars, while the vertical bars on the photometric points represent the 1$\sigma$ flux density error. The typical signal-to-noise of the NIRSpec and MIRI LRS spectra are $\sim$20 and $\sim$100 respectively. The spectrum shown in this figure is available as the Data behind the Figure.} \label{fig:speclabel}
\end{figure*}

\begin{figure}

\includegraphics[width=0.475\textwidth]{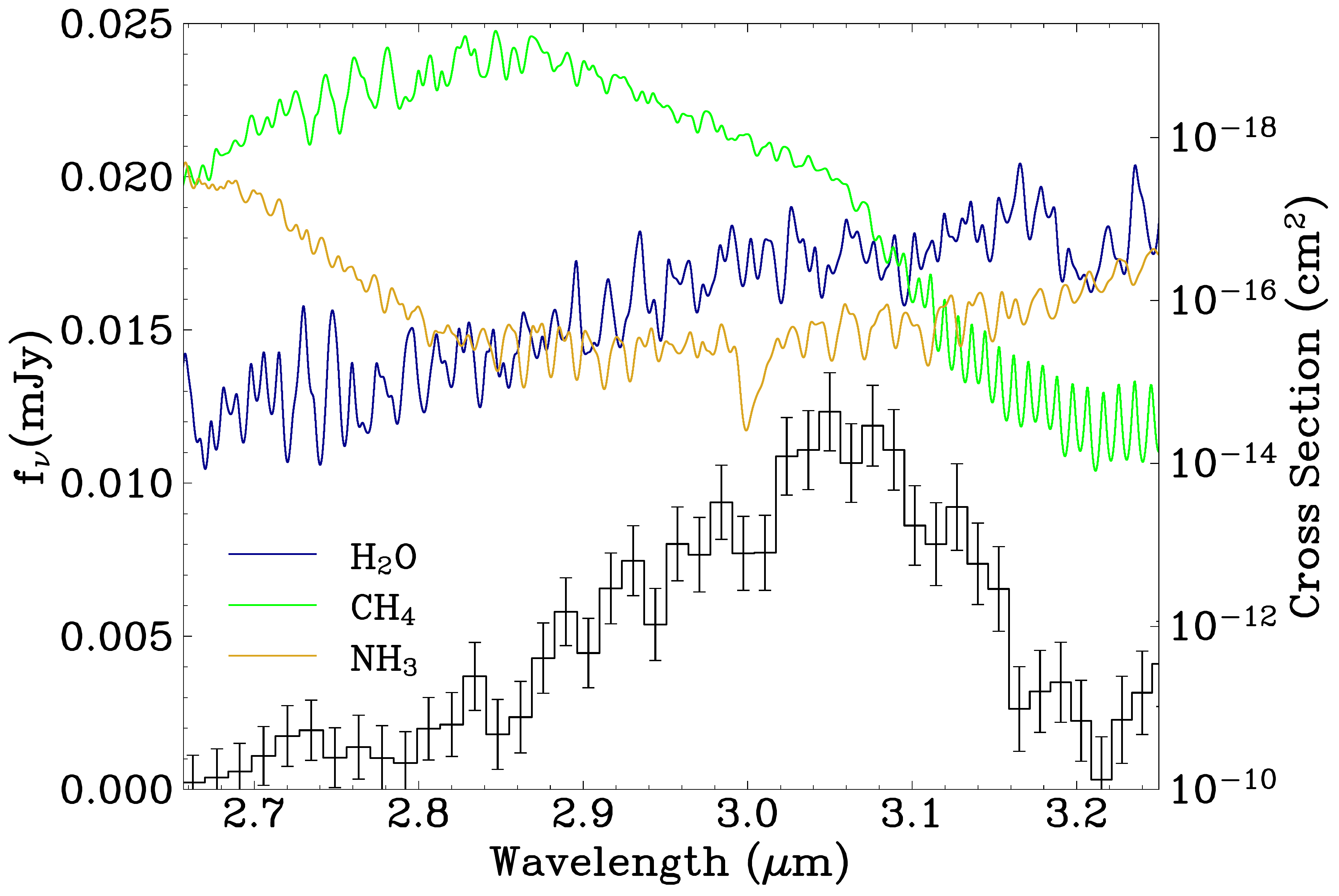}
\includegraphics[width=0.478\textwidth]{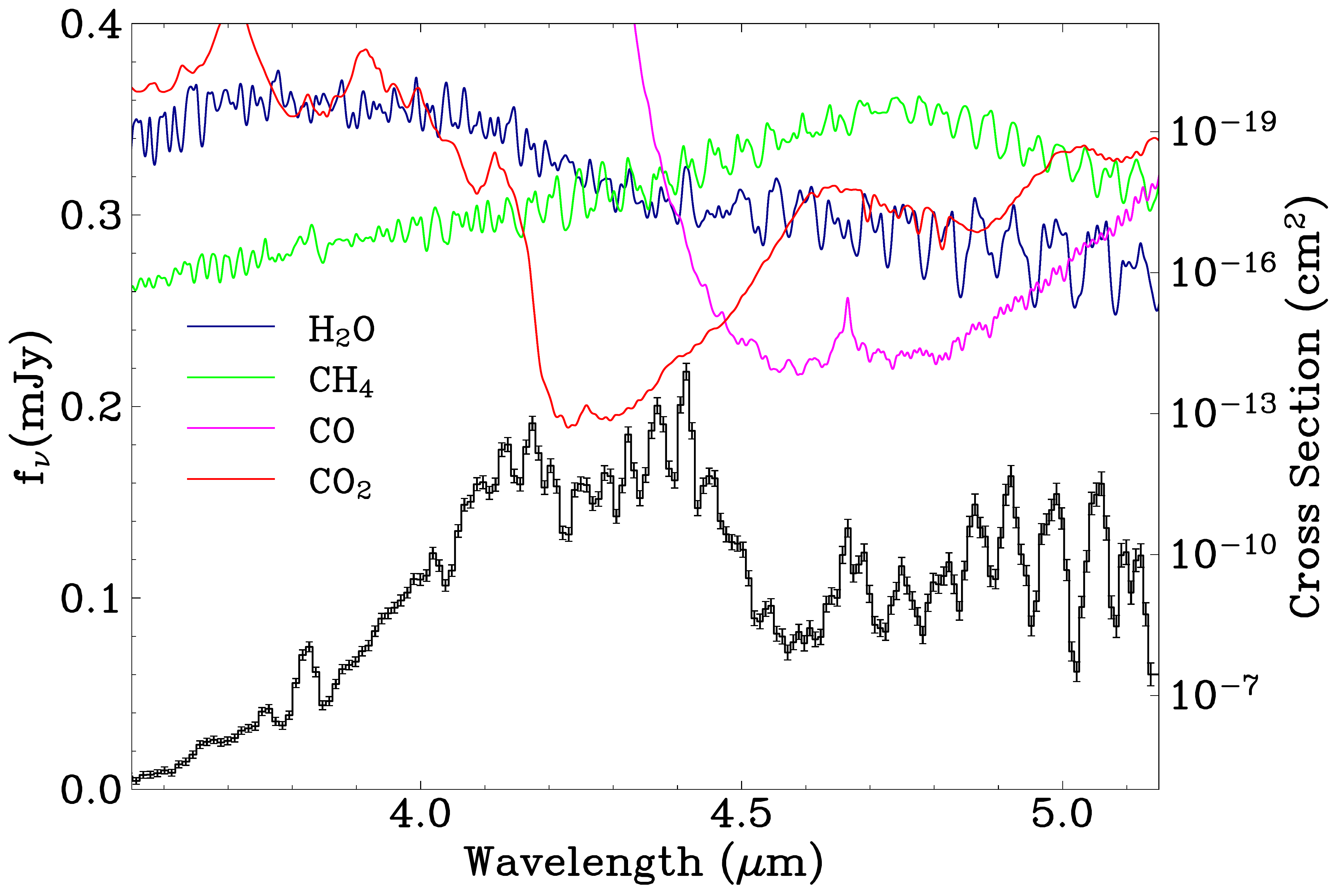}
\includegraphics[width=0.475\textwidth]{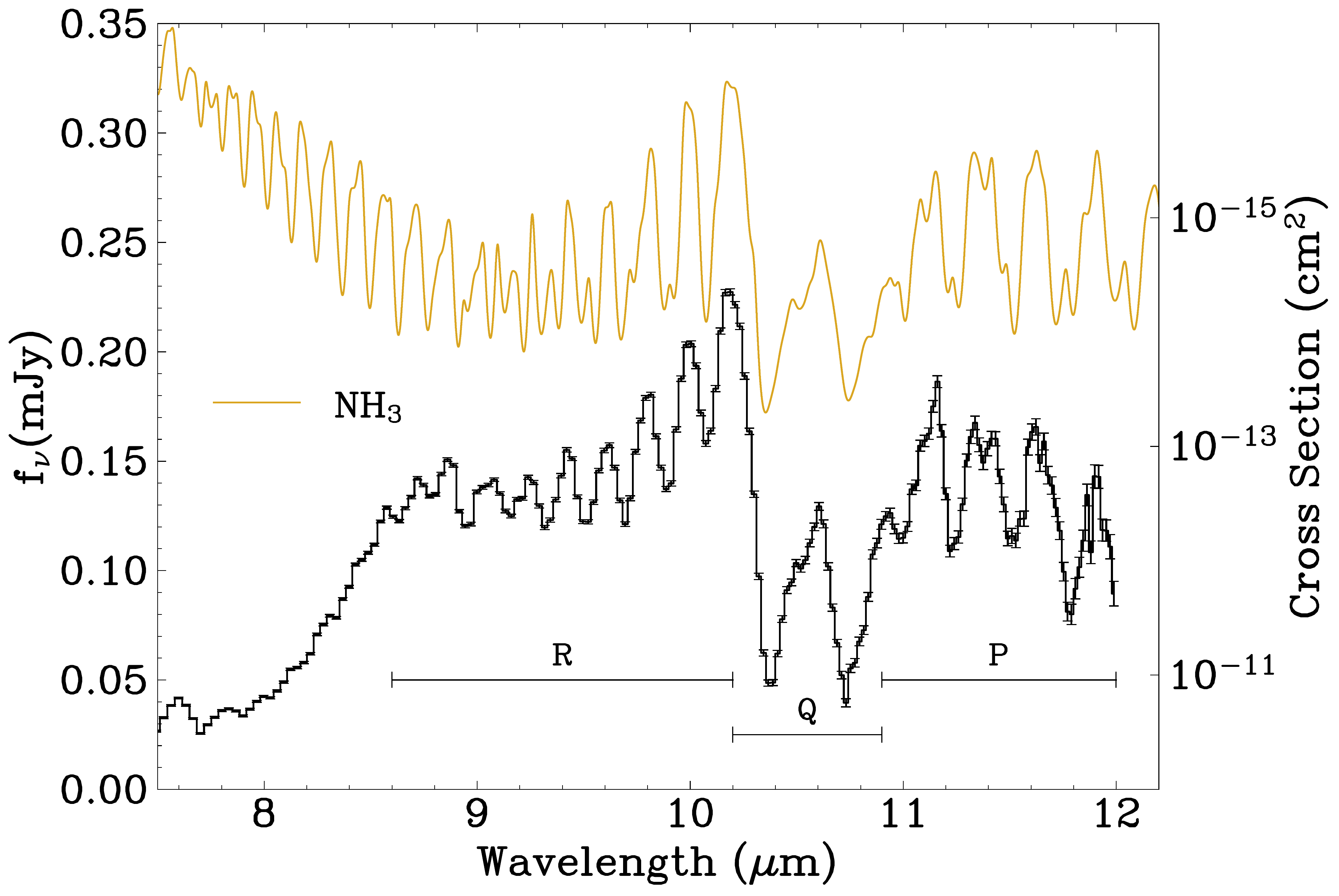}

\caption{Three sections of interest from our spectrum (black), with important molecular cross sections plotted over top (color). These cross sections were calculated at $P$ = 1 bar and  $T$ = 500 K. Note that the y-axis for the cross sections is inverted for clarity.
Top: We tentatively identify a new NH$_3$ absorption feature found at 3 \mic{}. H$_2$O and CH$_4$ set the red and blue edge of this peak, but the strong $Q$ branch of the $\nu_\mathrm{1}$ ammonia band carves out a small notch in the peak. Center: The 5 \mic{} peak is shaped by both CO and CO$_2$ features. Both of these molecules are not expected in equilibrium models of $\sim$500 K, but are found in models that account for vertical mixing in the atmosphere. Bottom: From 8--12 \mic{} ammonia is the dominant absorber in \wname's atmosphere. The $\nu_\mathrm{2}$ $Q$-branch doublet at 10.3 and 10.7 \mic{} has been seen before, but the $P$ and $R$ branches of ammonia on either side are seen for the first time.}
\label{fig:opacities}
\end{figure}

\section{Prominent Absorption Bands}\label{sec:spec}
Numerous fundamental, overtone, and combination rotational-vibrational bands of water ($\mathrm{H}_2\mathrm{O}$), methane ($\mathrm{CH}_4$), ammonia ($\mathrm{NH}_3$), carbon monoxide ($\mathrm{CO}$), and carbon dioxide ($\mathrm{CO}_2$) are present in the spectrum of \wname. While many of these bands were seen in the previously published spectra of \wname{} and warmer T dwarfs \citep[e.g.][]{oppenheimer_infrared_1995,oppenheimer_spectrum_1998,roellig_spitzer_2004,yamamura_akari_2010,schneider_hubble_2015}, this is the first time they can all be studied in a single spectrum. Figure \ref{fig:speclabel} shows the merged spectrum with key absorption features labeled, while Figure \ref{fig:opacities} shows the cross sections of these molecules at key portions of the spectrum. The cross sections were calculated at $P$ = 1 bar, $T$ = 500~K.} Other representative cross section plots can also be found in the literature \citep[e.g.][]{cushing_discovery_2011,morley_water_2014}.

Water and methane bands dominate large portions of the \wname{} spectrum. The water bands are at 1.05--1.2, 1.4--1.5, 1.8--2, 2.4--2.8, and 4.8--7.5 \mic{} and the methane bands are at 1.1--1.2, 1.3--1.5, 1.6--1.8, 3--4, and 6.8--8.3 \mic, all of which can be seen labeled in Figure \ref{fig:speclabel}. 

While methane is the main reservoir of carbon at \wname's effective temperatures of $\sim$500 K \citep{lodders_atmospheric_2002,kirkpatrick_field_2021}, there is still enough carbon monoxide and carbon dioxide in the atmosphere to see the $\nu_\mathrm{3}$ band of CO from 4.5--5.0 \mic{} and the $\nu_\mathrm{3}$ band of CO$_2$ at 4.2--4.35 \mic{} (center panel of Figure \ref{fig:opacities}). Both the CO$_2$ and CO bands appear slightly weaker by eye than those in the spectra of late T dwarfs presented by \citet{yamamura_akari_2010}. This is expected given the effective temperature of \wname{} is $\sim300$ K cooler.

Ammonia has been notoriously difficult to observe in the near-infrared, as it is blended with water and methane bands, as seen in Figure \ref{fig:speclabel}, and similarly water obscures the ammonia band at 5.5--7.1 \mic{}. Only at longer wavelengths (8.5--12 \mic) does ammonia absorption dominate the spectrum. We see the strong $\nu_\mathrm{2}$ $Q$-branch doublet at 10.5 \mic{}, and \wname's cooler effective temperature and \textit{JWST}'s sensitivity allows the $P$ and $R$ branches of ammonia on either side of the $Q$ branch to be clearly observed (bottom panel of Figure \ref{fig:opacities}).

We also tentatively identify a new ammonia feature at 3 \mic{}, presented in the top panel of Figure \ref{fig:opacities}. In the atmospheric window between the water and methane absorption bands, there is a small absorption feature which lines up with the $Q$ branch of the $\nu_\mathrm{1}$ band of ammonia. This portion of the spectrum has not previously been studied in Y dwarfs and provides another ammonia feature to constrain its abundance. While it is possible this is an artifact due to the signal-to-noise ratio in this portion of the spectrum being $\sim$10, the two data points that constitute the feature have flux density values more than 1$\sigma$ below the continuum on either side. 

\section{Bolometric Luminosity and Effective Temperature}\label{sec:prop}
The broad wavelength coverage provided by \textit{JWST} allows us to collect a substantial fraction of the light emitted by \wname{} and calculate the bolometric flux, $F_\mathrm{bol} = \int_0^\infty f_\lambda \, d\lambda$. We construct a complete spectral energy distribution by first linearly interpolating from zero flux at zero wavelength to the first data point of the combined spectrum at 0.96 \mic{} to capture the minutia of flux emitted at these short wavelengths. The combined spectrum then extends out to 12 \mic, and from there we linearly interpolate from the last data point in the spectrum through each of the 3 photometric filters at 15.065, 17.987, 20.795 \mic{}. Longward of the last photometric point, we approximate the spectral energy distribution as a Rayleigh-Jeans tail,
\begin{equation}
    f_{\lambda} = \frac{2ck_\mathrm{B}T}{\lambda^4} = \frac{C_\mathrm{RJ}}{\lambda^4},
\end{equation}
where $c$ is the speed of light, $k_\mathrm{B}$ is the Boltzmann constant, and $T$ is temperature. $C_\mathrm{RJ}$ is determined using the flux density at 20.795 \mic{} ($f_{\nu}=0.0731$ mJy).

To find \fbol{} we integrate under this spectral energy distribution. For the linearly interpolated regions and the Rayleigh-Jean tail, the integral is calculated analytically, while the spectral portion is integrated numerically using Simpson's rule. To estimate the uncertainty in \fbol{} we generate a million spectral energy distributions by randomly sampling from the distribution of each photometric and spectral data point. This includes randomly sampling the absolute calibration scaling factors derived from the \textit{Spitzer} [4.5] and F1000W photometry so the NIRSpec and MIRI LRS scaling are varied independently. From each of these million spectral energy distributions we compute an \fbol{} value. The mean and standard deviation of these values is \fbol{} =  $6.894 \pm 0.164\times10^{-17} \mathrm{\ W} \mathrm{\ m}^{-2}$, with only 5\% of the flux coming longward of 21 \mic{} and $<0.1$\% shortward of 0.96 \mic. \citet{dupuy_distances_2013} had previously calculated an \fbol{} for \wname{} of $4.94\times10^{-17} \mathrm{\ W} \mathrm{\ m}^{-2}$ using a near-infrared spectrum, \textit{Spitzer} photometry, and atmospheric models. This under-prediction shows the necessity of collecting the full spectral energy distribution to calculate an accurate \fbol.

\wname{} has a parallax of $73.6\pm2.0$ mas \citep[d=$13.57\pm0.37$ pc,][]{kirkpatrick_field_2021}, which allows us to compute \lbol{} as $L_\mathrm{bol} = 4\pi d^2F_\mathrm{bol}$. This results in \lbol{}$= 1.523 \pm 0.090\times10^{20} \,\mathrm{W}$ or $\mathrm{log}(L/\mathcal{L}_\odot^N)=-6.400\pm 0.025$, where $\mathcal{L}_\odot^N$ is the nominal solar luminosity of 3.828$\times 10^{26}$ W \citep{mamajek_iau_2015}. The bolometric flux and luminosity were used to calculate the apparent and absolute bolometric magnitudes using the zero-points from \citet{mamajek_iau_2015}, and are listed with the other parameters in Table \ref{tbl:fund}.

With \lbol, we could calculate the effective temperature if the radius were known since,
\begin{equation} \label{eqn:teff}
    T_\mathrm{eff} = \left( \frac{L_\mathrm{bol}}{4 \pi \sigma R^2} \right)^{\frac{1}{4}}.
\end{equation}
A direct measurement of \wname's radius is currently unfeasible, but due to the competing effects of Coulomb repulsion and electron degeneracy, the radii of brown dwarfs are $\sim$1 $\mathcal{R}_\mathrm{J}^\mathrm{N}$—the nominal value for
Jupiter’s equatorial radius of $7.1492 \times 10^7$ m \citep{mamajek_iau_2015}—across a wide range of ages and masses \citep[e.g.][]{burrows_theory_2001}. We adopt this radius to give a nominal effective temperature of $T_\mathrm{eff}^{R_\mathrm{Jup}}=452$ K.

We make a more rigorous estimate of the radius using the technique of \citet{saumon_molecular_2000}, where evolutionary models are used in combination with an observed \lbol, and an assumed or known age estimate. We use the Sonora Bobcat models \citep{marley_sonora_2021}, a new generation of self-consistent atmospheric and evolutionary models that include updated opacities and atmospheric chemistry. The evolutionary portion of these models include grids that predict how fundamental properties such as mass, radius, and luminosity evolve over time. We randomly draw a million ages and luminosities and interpolate across the models to generate a distribution of possible radii. The luminosities were drawn from our generated \lbol{} distribution, but we have no measurement for the age of \wname{} since it is difficult to estimate the age of an isolated brown dwarf. We chose a conservative uniform age distribution of 1--10 Gyr, which was selected based on the simulated age distributions of \citet{kirkpatrick_field_2021}, where the $<$600 K objects have a nearly uniform distribution older than 1~Gyr. We use Eq. \ref{eqn:teff} to calculate a \teff{} distribution from this randomly generated distribution of radii.

\begin{figure*}
\includegraphics[width=\textwidth]{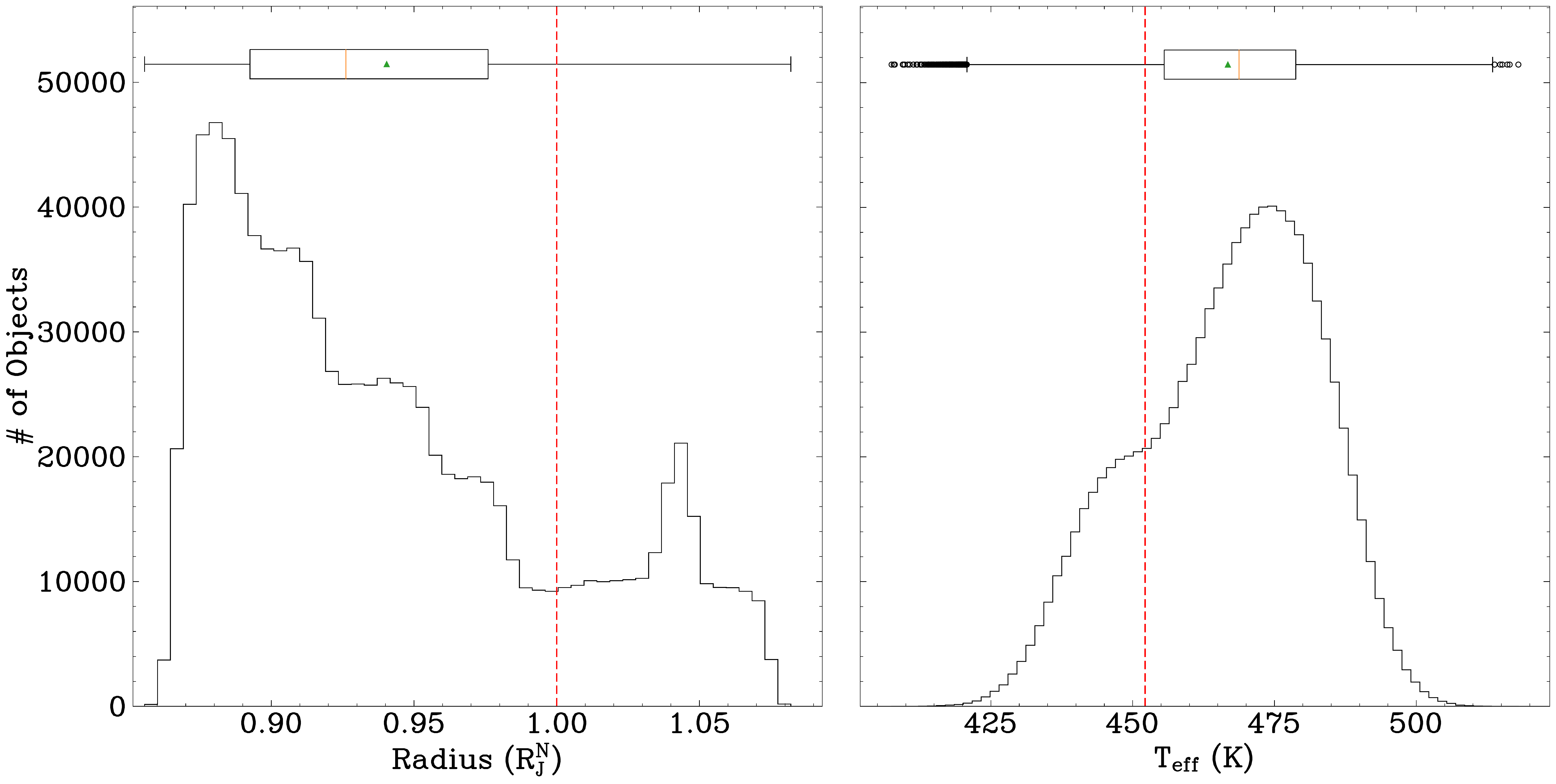}
\centering
\caption{The distributions of a million radii and effective temperatures for \wname{} drawn and calculated as described in Section \ref{sec:prop}, shown both as histograms and as box-and-whisker plots. The box-and-whisker plots divide the data into quartiles, with the median value marked as an orange line and the mean as a green triangle. We use luminosities drawn from a Gaussian distribution based on our calculated value along with ages drawn from a uniform 1--10 Gyr distribution to generate the radii distribution on the left. Each luminosity draw and corresponding radius is used to calculate a \teff, resulting in the distribution on the right. For reference we include a zeroth order approximation in the form of red lines at $1$ \rjup{} and $T_\mathrm{eff}^{R_\mathrm{J}}$. The pile-up at 1.04 \rjup{} in the model radii distribution translates to the bump in the \teff{} distribution around 445 K. }\label{fig:TRDist}
\end{figure*}

The resulting radii and \teff{} distributions are plotted in Figure \ref{fig:TRDist}, with box-and-whisker plots marking the quartiles. For reference we also include vertical lines indicating 1 \rjup{} and $T_\mathrm{eff}^{R_\mathrm{Jup}}$. Both distributions are asymmetric with tails extending to larger radii and cooler temperatures. We report the mean, median, and the interquartile values for each distribution in Table \ref{tbl:fund}, and we adopt our mean \teff=$467_{-18}^{+16}$ as the \teff{} of \wname. The 1 \rjup{} and $T_\mathrm{eff}^{R_\mathrm{Jup}}$ are both within $1\sigma$ of our semi-empirical values, at 6\% greater than our mean $R_\mathrm{mod}$=$0.940_{-0.057}^{+0.074}$, and 3\% lower than our mean \teff=$467_{-18}^{+16}$.

\begin{deluxetable}{lr}
\tablecaption{Physical Parameters of \wname \label{tbl:fund}}
\tablehead{
\colhead{Parameter} &
\colhead{Value}
}
\startdata
\fbol{} (W m$^{-2}$) & $6.894\pm0.164\times10^{-17}$\\
\lbol{} (W) & $1.523\pm0.090\times10^{20}$\\
$\mathrm{log}(L/\mathcal{L}_\odot^N)$ & $-6.400\pm0.025$ \\
$m_{\mathrm{bol}}$ (mag)& $21.406\pm0.026$\\
$M_{\mathrm{bol}}$ (mag)& $20.741\pm0.063$\\
$T_\mathrm{eff}^{R_\mathrm{J}}$ (K) & $452$ \\
\hline
\multicolumn{2}{c}{Uniform 1--10 Gyr Age Distribution}\\
\hline
Model Radius ($R_\mathrm{mod}$) Mean (\rjup) &$0.940_{-0.057}^{+0.074}$\tablenotemark{a} \\
$R_\mathrm{mod}$ Median (\rjup) & $0.926$ (IQR 0.893--0.976) \\
\teff{} Mean (K) &$467_{-18}^{+16}$ \tablenotemark{a} \\
\teff{} Median (K) & $469$ (IQR 456--479) \\
\enddata
\tablenotetext{a}{Reported 1$\sigma$ values enclose $\pm$34.134$\%$ of the distribution from the mean.}
\end{deluxetable}

Previously, the most precise effective temperature estimate for \wname{} was 436$\pm$88 K from \citet{kirkpatrick_field_2021}. These authors calculated the relation between $M_\mathrm{H}$ and \teff{} to make this estimate, as \citet{filippazzo_fundamental_2015} had found that for M, L, and T dwarfs this \teff{} relation has the smallest scatter. However, the scatter of the relation, the disparate sources and methods used to find the effective temperatures in the relation, and the $H$-band magnitude uncertainties create large errors on the order of $\sim$20\%. By calculating \teff{} using the bolometric flux from a broad wavelength spectral energy distribution, we are able to reduce the error down to $467_{-18}^{+16}$ K or $\sim$4\%. The dominant term in our error comes from our radius estimation, and if we knew our radius to infinite precision, our \teff{} error would drop to $\pm7$ K. This differs from the distance dominated errors of \citet{filippazzo_fundamental_2015}. Our precise distances are thanks to the astrometric work done by \citet{kirkpatrick_field_2021}; these and other astrometric measurements are foundational to our understanding of brown dwarfs and their physical parameters. 

\section{Model Fitting}\label{sec:fits}
We fit \wname's spectrum and photometry with the Sonora family of models \citep{marley_sonora_2021}, with the goals of 1) comparing our semi-empirical effective temperature estimates to the effective temperature of the best-fitting model, and 2) exploring the sensitivity of our observations to variations in metallicity, gravity, and disequilibrium chemistry. We searched for a best-fit model across both the Sonora Bobcat grid and a custom model grid generated by one of us \citep[S.M.,][]{mukherjee_picaso_2023, batalha_exoplanet_2019}. This custom grid is an extension and improvement of the Sonora Cholla models \citep{karalidi_sonora_2021} that will be published in the future, and includes an additional parameter \kzz, the vertical eddy diffusion coefficient. This parameter measures the vigor of atmospheric mixing in the vertical direction; a larger value of \kzz{} corresponds to a shorter mixing timescale. In our models, non-zero values of \kzz{} mean the abundances of CO/CH$_{4}$ and N$_2$/NH$_3$ are not in chemical equilibrium resulting in weaker or stronger absorption bands relative to equilibrium chemistry models.

For the Sonora Bobcat models, we explore a range of effective temperatures from 300--600 K at 25 K intervals, surface gravities (\logg) from 3--5.5 [$\mathrm{cm~s}^{-2}$] at 0.25 dex intervals, and metallicities ([M/H]) of $-0.5$, 0, and +0.5. The custom grid of models cover the same range of temperatures and gravities, but have metallicities of $-0.5$,$-0.3$, and 0, and additionally include the eddy diffusion coefficient (log \kzz{}) with values of 2, 4, 7, 8, and 9 [$\mathrm{cm}^2\ \mathrm{s}^{-1}$]. The \kzz{} values are constant with altitude, an unrealistic assuption but one that facilitates the calculation of grids such as this. Both models are cloudless.

We convolve the models at each data point with a Gaussian kernel to match the observed resolving power. Nominally, both the NIRSpec and MIRI LRS observations have a resolving power of $R\sim100$ but the actual resolving powers change by an order of magnitude across the two wavelength ranges. We assume that the NIRSpec observations are slit limited because the slit width is 0$\farcs$2 and the FWHM of a star is 0$\farcs$15 at $\sim$5 \mic. In this case, the resolving power is given by,
\begin{equation}
    R = \frac{\lambda}{2\delta\lambda},
\end{equation} 
where $\delta\lambda$ is the width of a pixel and the factor of two in the denominator is because the 0.$\farcs$2-wide slit subtends two pixels.  These results are consistent with pre-launch predictions\footnote{https://jwst-docs.stsci.edu/jwst-near-infrared-spectrograph/nirspec-instrumentation/nirspec-dispersers-and-filters}. We assume the MIRI LRS observations are source limited since the slit width is 0$\farcs$51 and the FWHM of a star is 0$\farcs$3 at $\sim$10 \mic. In this case, the resolving power is given by,
\begin{equation}
    R = \frac{\lambda}{\delta\lambda[\theta''(\lambda)/\theta_{\mathrm{pix}}'']},
\end{equation}
where $\theta''(\lambda)$ is the FWHM of a diffraction-limited star in arcseconds given by (64800/$\pi$)1.028$\lambda/D$ ($D$ = 6.5 m is the size of the \textit{JWST} primary mirror) and $\theta_\mathrm{pix}$ = 0$\farcs$11 is the angular size of a pixel. This also results in $R(\lambda)$ that is consistent with pre-launch predictions \citep{kendrew_mid-infrared_2015}.

For the fit to the convolved models, we include the three photometric points not used for flux calibration (F1500W, F1800W, and F2100W) by comparing them to synthetic photometry of each model calculated using the equation
\begin{equation}\label{eqn:synth}
    \langle f_\lambda \rangle= \frac{\int f_\lambda R_\lambda\lambda d\lambda}{\int R_\lambda\lambda d\lambda},
\end{equation}
from \citet{gordon_james_2022} where $f_\lambda$ is the model flux density and $R_\lambda$ is the bandpass function. The MIRI bandpasses were taken from the Spanish Virual Observatory (SVO)\footnote{http://svo2.cab.inta-csic.es/svo/theory/fps3/index.php}.

The best-fit model in the grid is the one that minimizes $\chi^2$, defined as:
\begin{equation}
    \chi^2 = \sum_i \left( \frac{f_{\lambda,i}-M_{\lambda,i} C}{\sigma_i^2} \right) ^2,
\end{equation}
where $M_{\lambda,i}$, $f_{\lambda,i}$, and $\sigma_{\lambda,i}$ are respectively the model flux density, observed flux density, and the observed flux density uncertainty in Jansky at each data point $i$. $C$ is the scaling factor for the model spectrum that minimizes $\chi^2$, defined by:
\begin{equation}\label{eqn:C}
    C=\frac{\sum\limits_{i}M_{\lambda,i} f_{\lambda,i}/\sigma_{i}^2}{\sum\limits_{i}{M_{\lambda,i}^2/\sigma_{i}^2}}.
\end{equation}

\begin{figure*}
\includegraphics[width=\textwidth]{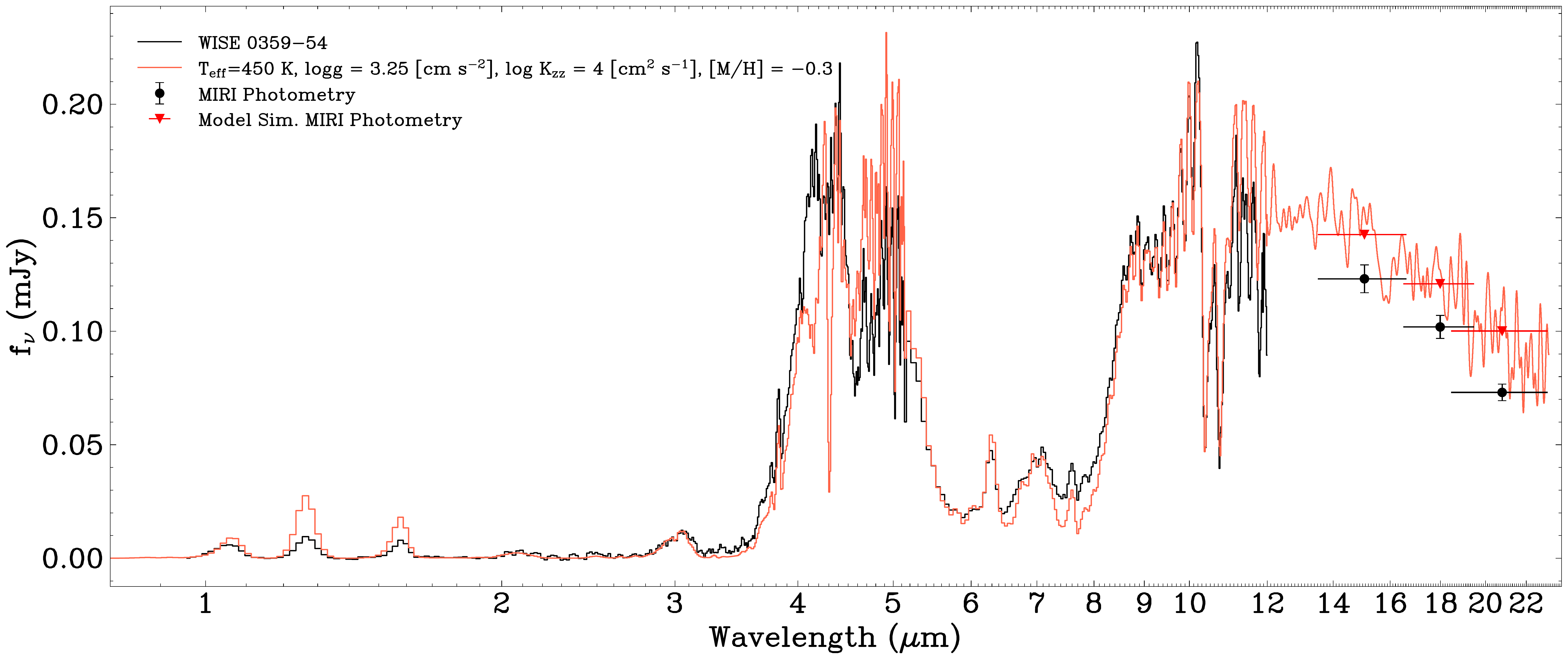}
\centering
\caption{The best-fitting Sonora model and its simulated photometry (orange) for our observed spectrum and photometry (black) with wavelength in log space. Overall, the fit is good given the broad wavelength coverage of the observations, with the model having some difficulties matching the features at the 5 \mic{} peak and an over-prediction of flux past 11 \mic. The effective temperature of the model is within 1$\sigma$ to our calculated effective temperature.}
\label{fig:modelfit}
\end{figure*}

The best-fitting model is from the custom grid and has \teff{} = 450 K, \logg{} = 3.25 [$\mathrm{cm~s}^{-2}$], log \kzz{} = 4 [$\mathrm{cm}^2\ \mathrm{s}^{-1}$], [M/H] = $-0.3$ with $\chi^2=77953$ (d.o.f. = 5). This model fit is shown in Figure \ref{fig:modelfit}. While statistically the fit is poor given the $\chi^2$, the overall fit is relatively good given the broad wavelength coverage of the observations. 

There are, however, several mismatches that deserve mention. We see an excellent fit of the ammonia features shortward of 11 \mic{}, but from 11--21 \mic{} the model flux density is $\sim$ 15$\%$ higher than the data. There are also mismatches within the 5 \mic{} peak. The model under-predicts the flux emitted from 4--4.5 \mic{} as a result of PH$_3$ absorption that is not seen in our observations, and the CO band from 4.5--5.4 \mic{} is not deep enough in the model spectrum. Another notable divergence occurs in the near-infrared ($<$3 \mic), where the model predicts almost double the observed flux density in the $J$ and $H$ bands. This portion of the spectrum is where most Y dwarf observations have been carried out, including a \textit{Hubble} WFC3 spectrum (0.9--1.1 \mic) of \wname{} from \citet{schneider_hubble_2015}. They used this spectrum to derive a \teff{} of 400 K, lower than both our best-fit model and our semi-empirical \teff. This spectral mismatch and \teff{} discrepancy underscores the difficulty in understanding these cool objects with only near-infrared spectra. 

Our best-fit effective temperature of 450 K is close to our empirically derived temperature of $467_{-18}^{+16}$ K. There is a lack of precision in the effective temperature preferred by the model, as the 10 best-fitting models include an equal distribution of 425, 450, and 475 K temperatures. That said, considering this is the first ever comparison of a complete model to an observed spectral energy distribution in this temperature range the alignment in \teff{} is remarkable. We also note that hotter models fit the spectrum and photometry better past 11 \mic{} as the peak of the Planck function shifts to shorter wavelengths, and heavier weighting of these wavelengths or an extended spectrum would likely result in a higher best-fit effective temperature. 

We find a strong preference for models with a disequilibrium chemistry, particularly log~\kzz{} $\geq$ 4, with only 3 of the top 100 best-fitting models having a log~$\kzz{}=2$ and none with log~$\kzz{}=0$. This is because disequilibrium chemistry is required to match the observed depths of the CO and NH$_3$ bands centered at 4.7 and 11.5 \mic{}. However, non-zero \kzz{} values also result in deep phosphine bands at 4.15 and 4.3 \mic{} in the model spectra that is not observed in our spectrum. \citet{miles_observations_2020} also did not detect phosphine in their $M$-band spectra of late T and Y dwarfs despite expecting to at the temperature and \kzz{} values needed to fit the 4.7 \mic{} CO band. One likely source of this issue is the unrealistic assumption that \kzz{} is constant with altitude \citep{mukherjee_probing_2022}. The marked lack of PH$_3$ may point to shortcomings in our understanding of phosphorus chemistry and reaction pathways under these atmospheric conditions \citep{visscher_atmospheric_2006,morley_l_2018}. Future and more complex models will be required to understand these chemical interactions. The presence of the phosphine band in the models has the added effect of masking the overlapping CO$_2$ band centered at 4.2 \mic{}, making it difficult for us to draw any conclusions about how this CO$_2$ band is fit by the model.

There is also a preference in the model fits for sub-solar metallicity ([M/H]=$-0.3$, 9 of the top 10) and for \logg{} $\leq$ 3.75 (46 of the top 50). Lower metallicity increases the flux at the blue edge of the 5 \mic{} peak as shown in \citet{cushing_improved_2021} and makes the 8.5--12 \mic{} ammonia band weaker, both of which lead to a better fit to our spectrum. In general, lower gravities slightly suppress the flux blueward of 5.5 \mic{}, and decrease absorption in the $P$ and $R$ branches of the 10.5 \mic{} ammonia feature.

The scaling factor $C$ calculated by Eq. \ref{eqn:C} has a physical analog of $R^2$/$d^2$, where $R$ is the radius of the brown dwarf and $d$ is the distance to it. Our best-fit model has $C$=3.48$\times10^{-20}$, which at a distance of 13.57$\pm$0.37 pc \citep{kirkpatrick_field_2021} gives a radius of 1.09 \rjup{}. This is a little over 2$\sigma$ larger than our mean simulated radius.

\begin{figure}
\includegraphics[width=.45\textwidth]{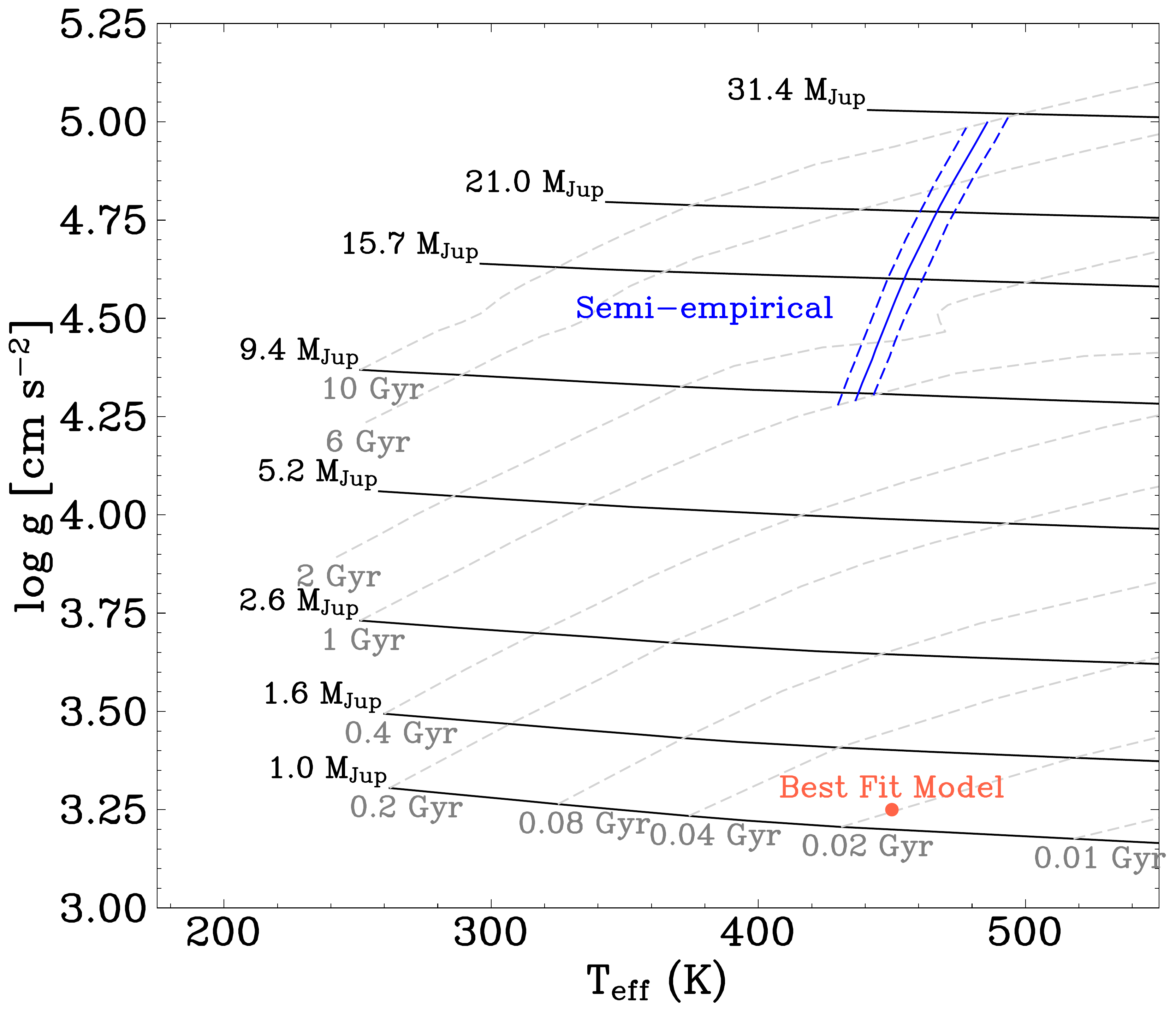}
\centering
\caption{Isochrones (grey) and cooling tracks (black) from the Sonora Bobcat evolutionary models in the effective temperature surface gravity plane. We plot the best-fit model at its \logg{} and \teff{} (orange), as well as the loci of points consistent with the observed bolometric luminosity and an age range estimate (1--10 Gyr, blue), with the blue dashed lines indicating $\pm1\sigma$. The best-fit model's low surface gravity corresponds to an age of 20 Myr and a significantly lower mass of $\sim$1 \mjup, compared to the lowest value of $\sim$10 \mjup{} of our semi-empirical range.}\label{fig:logg}
\end{figure}

We use the isochrones and cooling tracks from the Sonora Bobcat solar metallicity evolutionary models to estimate the mass of \wname{} for the fitted \logg{} and \teff{} values. In Figure \ref{fig:logg} we show the best-fit model values plotted in \teff{} and \logg{} space along with the isochrones and cooling tracks, as well as the range covered by our measured \lbol{} and age range estimate of 1--10 Gyr. The best-fit model corresponds to a mass of $\sim1$ \mjup{} at an age of 20 Myr, while our semi-empirical measurement corresponds to a mass range from $\sim$9--31 \mjup. These masses and ages are clearly not in agreement, driven by the low gravity of the best-fit model. The simulations of volume-limited populations of brown dwarfs with effective temperatures of 450--600 K show that the vast majority of them are old with a median age of 5 Gyr \citep{kirkpatrick_field_2021}. This suggests that caution should be taken when interpreting the surface gravity estimate of \wname.

\begin{acknowledgments}
This work is based [in part] on observations made with the NASA/ESA/CSA James Webb Space Telescope. The data were obtained from the Mikulski Archive for Space Telescopes at the Space Telescope Science Institute, which is operated by the Association of Universities for Research in Astronomy, Inc., under NASA contract NAS 5-03127 for JWST. These observations are associated with program \#2302. Support for program \#2302 was provided by NASA through a grant from the Space Telescope Science Institute, which is operated by the Association of Universities for Research in Astronomy, Inc., under NASA contract NAS 5-03127. This research has benefited from the Y Dwarf Compendium maintained by Michael Cushing at \url{https://sites.google.com/view/ydwarfcompendium}, as well as conversations with Rocio Kiman and Genaro Su\'{a}rez about the \textit{JWST} pipeline. This research has made use of the Spanish Virtual Observatory (https://svo.cab.inta-csic.es) project funded by MCIN/AEI/10.13039/501100011033/ through grant PID2020-112949GB-I00. The JWST data presented in this paper were obtained from the Mikulski Archive for Space Telescopes (MAST) at the Space Telescope Science Institute. The specific observations analyzed can be accessed via \dataset[10.17909/96qb-wh63]{https://doi.org/10.17909/96qb-wh63}.
\end{acknowledgments}

\facility{JWST (NIRSpec, MIRI LRS)}
\software{NumPy \citep{harris_array_2020}, smplotlib \citep{li_astrojacoblismplotlib_2023}}
\appendix
\section{Synthetic \textit{JWST} Magnitudes of WISE~0359--54}
We provide synthetic \textit{JWST} photometry for \wname{} in all filters that fall entirely within our observed wavelength range in Tables \ref{tbl:SynNIRCAM} and \ref{tbl:SynMIRI}. All values were calculated using Eq. \ref{eqn:synth} from \citet{gordon_james_2022}. 
\begin{deluxetable}{lrc}
\tablecaption{Synthetic \textit{JWST} NIRCam Photometry of \wname \label{tbl:SynNIRCAM}}
\tablehead{
\colhead{Filter Name} &
\colhead{$\lambda_{pivot}$} &
\colhead{$f_{\nu}$}
}
\startdata
F115W & 1.154 \mic  & 3.34e-06 Jy\\
F140M & 1.405 \mic  & 1.25e-07 Jy\\
F150W & 1.501 \mic  & 1.71e-06 Jy\\
F150W2 & 1.671 \mic  & 1.85e-06 Jy\\
F162M & 1.627 \mic  & 3.03e-06 Jy\\
F164N & 1.645 \mic  & 5.38e-07 Jy\\
F182M & 1.845 \mic  & 2.75e-07 Jy\\
F187N & 1.874 \mic  & 5.20e-08 Jy\\
F200W & 1.988 \mic  & 1.03e-06 Jy\\
F210M & 2.096 \mic  & 2.09e-06 Jy\\
F212N & 2.121 \mic  & 1.70e-06 Jy\\
F250M & 2.503 \mic  & 1.44e-06 Jy\\
F277W & 2.777 \mic  & 3.85e-06 Jy\\
F300M & 2.996 \mic  & 7.42e-06 Jy\\
F322W2 & 3.247 \mic  & 1.69e-05 Jy\\
F323N & 3.237 \mic  & 2.80e-06 Jy\\
F335M & 3.362 \mic  & 3.92e-06 Jy\\
F356W & 3.565 \mic  & 2.43e-05 Jy\\
F360M & 3.623 \mic  & 1.92e-05 Jy\\
F405N & 4.053 \mic  & 1.27e-04 Jy\\
F410M & 4.084 \mic  & 1.30e-04 Jy\\
F430M & 4.281 \mic  & 1.64e-04 Jy\\
F444W & 4.402 \mic  & 1.28e-04 Jy\\
F460M & 4.630 \mic  & 9.50e-05 Jy\\
F466N & 4.654 \mic  & 1.11e-04 Jy\\
F470N & 4.708 \mic  & 9.81e-05 Jy\\
F480M & 4.817 \mic  & 1.13e-04 Jy\\
\enddata
\end{deluxetable}

\begin{deluxetable}{lrc}
\tablecaption{Synthetic \textit{JWST} MIRI Photometry of \wname \label{tbl:SynMIRI}}
\tablehead{
\colhead{Filter Name} &
\colhead{$\lambda_{pivot}$} &
\colhead{$f_{\nu}$}
}
\startdata
F560W & 5.635 \mic  & 4.47e-05 Jy\\
F770W & 7.639 \mic  & 5.09e-05 Jy\\
F1065C & 10.563 \mic  & 9.61e-05 Jy\\
F1130W & 11.309 \mic  & 1.40e-04 Jy\\
F1140C & 11.310 \mic  & 1.42e-04 Jy\\
\enddata
\end{deluxetable}

\bibliography{sample631.bib}

\begin{thebibliography}{}
\expandafter\ifx\csname natexlab\endcsname\relax\def\natexlab#1{#1}\fi
\providecommand{\url}[1]{\href{#1}{#1}}
\providecommand{\dodoi}[1]{doi:~\href{http://doi.org/#1}{\nolinkurl{#1}}}
\providecommand{\doeprint}[1]{\href{http://ascl.net/#1}{\nolinkurl{http://ascl.net/#1}}}
\providecommand{\doarXiv}[1]{\href{https://arxiv.org/abs/#1}{\nolinkurl{https://arxiv.org/abs/#1}}}

\bibitem[{Batalha {et~al.}(2019)Batalha, Marley, Lewis, \&
  Fortney}]{batalha_exoplanet_2019}
Batalha, N.~E., Marley, M.~S., Lewis, N.~K., \& Fortney, J.~J. 2019, The
  Astrophysical Journal, 878, 70, \dodoi{10.3847/1538-4357/ab1b51}

\bibitem[{Brooks {et~al.}(2023)Brooks, Kirkpatrick, Meisner, Gelino, Gagliuffi,
  Marocco, Schneider, Faherty, Casewell, Raghu, Kuchner, Worlds, \&
  Collaboration}]{brooks_long-term_2023}
Brooks, H., Kirkpatrick, J.~D., Meisner, A.~M., {et~al.} 2023, Long-term
  4.6\${\textbackslash}mu\$m {Variability} in {Brown} {Dwarfs} and a {New}
  {Technique} for {Identifying} {Brown} {Dwarf} {Binary} {Candidates},  arXiv.
\newblock \url{http://arxiv.org/abs/2304.05630}

\bibitem[{Burrows {et~al.}(2001)Burrows, Hubbard, Lunine, \&
  Liebert}]{burrows_theory_2001}
Burrows, A., Hubbard, W.~B., Lunine, J.~I., \& Liebert, J. 2001, Reviews of
  Modern Physics, 73, 719, \dodoi{10.1103/RevModPhys.73.719}

\bibitem[{Cushing {et~al.}(2011)Cushing, Kirkpatrick, Gelino, Griffith,
  Skrutskie, Mainzer, Marsh, Beichman, Burgasser, Prato, Simcoe, Marley,
  Saumon, Freedman, Eisenhardt, \& Wright}]{cushing_discovery_2011}
Cushing, M.~C., Kirkpatrick, J.~D., Gelino, C.~R., {et~al.} 2011, The
  Astrophysical Journal, 743, 50, \dodoi{10.1088/0004-637X/743/1/50}

\bibitem[{Cushing {et~al.}(2021)Cushing, Schneider, Kirkpatrick, Morley,
  Marley, Gelino, Mace, Wright, Eisenhardt, Skrutskie, \&
  Marsh}]{cushing_improved_2021}
Cushing, M.~C., Schneider, A.~C., Kirkpatrick, J.~D., {et~al.} 2021, The
  Astrophysical Journal, 920, 20, \dodoi{10.3847/1538-4357/ac12cb}

\bibitem[{Dupuy \& Kraus(2013)}]{dupuy_distances_2013}
Dupuy, T.~J., \& Kraus, A.~L. 2013, Science, 341, 1492,
  \dodoi{10.1126/science.1241917}

\bibitem[{Filippazzo {et~al.}(2015)Filippazzo, Rice, Faherty, Cruz, Van~Gordon,
  \& Looper}]{filippazzo_fundamental_2015}
Filippazzo, J.~C., Rice, E.~L., Faherty, J., {et~al.} 2015, The Astrophysical
  Journal, 810, 158, \dodoi{10.1088/0004-637X/810/2/158}

\bibitem[{Gordon {et~al.}(2022)Gordon, Bohlin, Sloan, Rieke, Volk, Boyer,
  Muzerolle, Schlawin, Deustua, Hines, Kraemer, Mullally, \&
  Su}]{gordon_james_2022}
Gordon, K.~D., Bohlin, R., Sloan, G.~C., {et~al.} 2022, The Astronomical
  Journal, 163, 267, \dodoi{10.3847/1538-3881/ac66dc}

\bibitem[{Harris {et~al.}(2020)Harris, Millman, van~der Walt, Gommers,
  Virtanen, Cournapeau, Wieser, Taylor, Berg, Smith, Kern, Picus, Hoyer, van
  Kerkwijk, Brett, Haldane, del Río, Wiebe, Peterson, Gérard-Marchant,
  Sheppard, Reddy, Weckesser, Abbasi, Gohlke, \& Oliphant}]{harris_array_2020}
Harris, C.~R., Millman, K.~J., van~der Walt, S.~J., {et~al.} 2020, Nature, 585,
  357, \dodoi{10.1038/s41586-020-2649-2}

\bibitem[{Jakobsen {et~al.}(2022)Jakobsen, Ferruit, Alves~de Oliveira, Arribas,
  Bagnasco, Barho, Beck, Birkmann, Böker, Bunker, Charlot, de~Jong, de~Marchi,
  Ehrenwinkler, Falcolini, Fels, Franx, Franz, Funke, Giardino, Gnata, Holota,
  Honnen, Jensen, Jentsch, Johnson, Jollet, Karl, Kling, Köhler, Kolm, Kumari,
  Lander, Lemke, López-Caniego, Lützgendorf, Maiolino, Manjavacas, Marston,
  Maschmann, Maurer, Messerschmidt, Moseley, Mosner, Mott, Muzerolle, Pirzkal,
  Pittet, Plitzke, Posselt, Rapp, Rauscher, Rawle, Rix, Rödel, Rumler, Sabbi,
  Salvignol, Schmid, Sirianni, Smith, Strada, te~Plate, Valenti, Wettemann,
  Wiehe, Wiesmayer, Willott, Wright, Zeidler, \&
  Zincke}]{jakobsen_near-infrared_2022}
Jakobsen, P., Ferruit, P., Alves~de Oliveira, C., {et~al.} 2022, Astronomy \&
  Astrophysics, 661, A80, \dodoi{10.1051/0004-6361/202142663}

\bibitem[{Karalidi {et~al.}(2021)Karalidi, Marley, Fortney, Morley, Saumon,
  Lupu, Visscher, \& Freedman}]{karalidi_sonora_2021}
Karalidi, T., Marley, M., Fortney, J.~J., {et~al.} 2021, The Astrophysical
  Journal, 923, 269, \dodoi{10.3847/1538-4357/ac3140}

\bibitem[{Kendrew {et~al.}(2015)Kendrew, Scheithauer, Bouchet, Amiaux,
  Azzollini, Bouwman, Chen, Dubreuil, Fischer, Glasse, Greene, Lagage, Lahuis,
  Ronayette, Wright, \& Wright}]{kendrew_mid-infrared_2015}
Kendrew, S., Scheithauer, S., Bouchet, P., {et~al.} 2015, Publications of the
  Astronomical Society of the Pacific, 127, 000, \dodoi{10.1086/682255}

\bibitem[{Kirkpatrick {et~al.}(2012)Kirkpatrick, Gelino, Cushing, Mace,
  Griffith, Skrutskie, Marsh, Wright, Eisenhardt, McLean, Mainzer, Burgasser,
  Tinney, Parker, \& Salter}]{kirkpatrick_further_2012}
Kirkpatrick, J.~D., Gelino, C.~R., Cushing, M.~C., {et~al.} 2012, The
  Astrophysical Journal, 753, 156, \dodoi{10.1088/0004-637X/753/2/156}

\bibitem[{Kirkpatrick {et~al.}(2021)Kirkpatrick, Gelino, Faherty, Meisner,
  Caselden, Schneider, Marocco, Cayago, Smart, Eisenhardt, Kuchner, Wright,
  Cushing, Allers, Bardalez~Gagliuffi, Burgasser, Gagné, Logsdon, Martin,
  Ingalls, Lowrance, Abrahams, Aganze, Gerasimov, Gonzales, Hsu, Kamraj, Kiman,
  Rees, Theissen, Ammar, Andersen, Beaulieu, Colin, Elachi, Goodman, Gramaize,
  Hamlet, Hong, Jonkeren, Khalil, Martin, Pendrill, Pumphrey, Rothermich,
  Sainio, Stenner, Tanner, Thévenot, Voloshin, Walla, \&
  Wędracki}]{kirkpatrick_field_2021}
Kirkpatrick, J.~D., Gelino, C.~R., Faherty, J.~K., {et~al.} 2021, The
  Astrophysical Journal Supplement Series, 253, 7,
  \dodoi{10.3847/1538-4365/abd107}

\bibitem[{Leggett {et~al.}(2017)Leggett, Tremblin, Esplin, Luhman, \&
  Morley}]{leggett_y-type_2017}
Leggett, S.~K., Tremblin, P., Esplin, T.~L., Luhman, K.~L., \& Morley, C.~V.
  2017, The Astrophysical Journal, 842, 118, \dodoi{10.3847/1538-4357/aa6fb5}

\bibitem[{Li(2023)}]{li_astrojacoblismplotlib_2023}
Li, J. 2023, {AstroJacobLi}/smplotlib: v0.0.6,  Zenodo,
  \dodoi{10.5281/zenodo.7839250}

\bibitem[{Lodders \& Fegley(2002)}]{lodders_atmospheric_2002}
Lodders, K., \& Fegley, B. 2002, Icarus, 155, 393,
  \dodoi{10.1006/icar.2001.6740}

\bibitem[{Mamajek {et~al.}(2015)Mamajek, Torres, Prsa, Harmanec, Asplund,
  Bennett, Capitaine, Christensen-Dalsgaard, Depagne, Folkner, Haberreiter,
  Hekker, Hilton, Kostov, Kurtz, Laskar, Mason, Milone, Montgomery, Richards,
  Schou, \& Stewart}]{mamajek_iau_2015}
Mamajek, E.~E., Torres, G., Prsa, A., {et~al.} 2015, {IAU} 2015 {Resolution}
  {B2} on {Recommended} {Zero} {Points} for the {Absolute} and {Apparent}
  {Bolometric} {Magnitude} {Scales},  arXiv.
\newblock \url{http://arxiv.org/abs/1510.06262}

\bibitem[{Marley {et~al.}(2021)Marley, Saumon, Visscher, Lupu, Freedman,
  Morley, Fortney, Seay, Smith, Teal, \& Wang}]{marley_sonora_2021}
Marley, M.~S., Saumon, D., Visscher, C., {et~al.} 2021, The Astrophysical
  Journal, 920, 85, \dodoi{10.3847/1538-4357/ac141d}

\bibitem[{Miles {et~al.}(2020)Miles, Skemer, Morley, Marley, Fortney, Allers,
  Faherty, Geballe, Visscher, Schneider, Lupu, Freedman, \&
  Bjoraker}]{miles_observations_2020}
Miles, B.~E., Skemer, A. J.~I., Morley, C.~V., {et~al.} 2020, The Astronomical
  Journal, 160, 63, \dodoi{10.3847/1538-3881/ab9114}

\bibitem[{Morley {et~al.}(2014)Morley, Marley, Fortney, Lupu, Saumon, Greene,
  \& Lodders}]{morley_water_2014}
Morley, C.~V., Marley, M.~S., Fortney, J.~J., {et~al.} 2014, The Astrophysical
  Journal, 787, 78, \dodoi{10.1088/0004-637X/787/1/78}

\bibitem[{Morley {et~al.}(2018)Morley, Skemer, Allers, Marley, Faherty,
  Visscher, Beiler, Miles, Lupu, Freedman, Fortney, Geballe, \&
  Bjoraker}]{morley_l_2018}
Morley, C.~V., Skemer, A.~J., Allers, K.~N., {et~al.} 2018, The Astrophysical
  Journal, 858, 97, \dodoi{10.3847/1538-4357/aabe8b}

\bibitem[{Mukherjee {et~al.}(2023)Mukherjee, Batalha, Fortney, \&
  Marley}]{mukherjee_picaso_2023}
Mukherjee, S., Batalha, N.~E., Fortney, J.~J., \& Marley, M.~S. 2023, The
  Astrophysical Journal, 942, 71, \dodoi{10.3847/1538-4357/ac9f48}

\bibitem[{Mukherjee {et~al.}(2022)Mukherjee, Fortney, Batalha, Karalidi,
  Marley, Visscher, Miles, \& Skemer}]{mukherjee_probing_2022}
Mukherjee, S., Fortney, J.~J., Batalha, N.~E., {et~al.} 2022, The Astrophysical
  Journal, 938, 107, \dodoi{10.3847/1538-4357/ac8dfb}

\bibitem[{Murakami {et~al.}(2007)Murakami, Baba, Barthel, Clements, Cohen, Doi,
  Enya, Figueredo, Fujishiro, Fujiwara, Fujiwara, Garcia-Lario, Goto, Hasegawa,
  Hibi, Hirao, Hiromoto, Hong, Imai, Ishigaki, Ishiguro, Ishihara, Ita, Jeong,
  Jeong, Kaneda, Kataza, Kawada, Kawai, Kawamura, Kessler, Kester, Kii, Kim,
  Kim, Kobayashi, Koo, Kwon, Lee, Lorente, Makiuti, Matsuhara, Matsumoto,
  Matsuo, Matsuura, MÜller, Murakami, Nagata, Nakagawa, Naoi, Narita, Noda,
  Oh, Ohnishi, Ohyama, Okada, Okuda, Oliver, Onaka, Ootsubo, Oyabu, Pak, Park,
  Pearson, Rowan-Robinson, Saito, Sakon, Salama, Sato, Savage, Serjeant,
  Shibai, Shirahata, Sohn, Suzuki, Takagi, Takahashi, TanabÉ, Takeuchi,
  Takita, Thomson, Uemizu, Ueno, Usui, Verdugo, Wada, Wang, Watabe, Watarai,
  White, Yamamura, Yamauchi, \& Yasuda}]{murakami_infrared_2007}
Murakami, H., Baba, H., Barthel, P., {et~al.} 2007, Publications of the
  Astronomical Society of Japan, 59, S369, \dodoi{10.1093/pasj/59.sp2.S369}

\bibitem[{Oppenheimer {et~al.}(1998)Oppenheimer, Kulkarni, Matthews, \&
  Kerkwijk}]{oppenheimer_spectrum_1998}
Oppenheimer, B.~R., Kulkarni, S.~R., Matthews, K., \& Kerkwijk, M. H.~v. 1998,
  The Astrophysical Journal, 502, 932, \dodoi{10.1086/305928}

\bibitem[{Oppenheimer {et~al.}(1995)Oppenheimer, Kulkarni, Matthews, \&
  Nakajima}]{oppenheimer_infrared_1995}
Oppenheimer, B.~R., Kulkarni, S.~R., Matthews, K., \& Nakajima, T. 1995,
  Science (New York, N.Y.), 270, 1478, \dodoi{10.1126/science.270.5241.1478}

\bibitem[{Reach {et~al.}(2005)Reach, Megeath, Cohen, Hora, Carey, Surace,
  Willner, Barmby, Wilson, Glaccum, Lowrance, Marengo, \&
  Fazio}]{reach_absolute_2005}
Reach, W., Megeath, S., Cohen, M., {et~al.} 2005, Publications of the
  Astronomical Society of the Pacific, 117, 978, \dodoi{10.1086/432670}

\bibitem[{Rieke {et~al.}(2015)Rieke, Wright, Böker, Bouwman, Colina, Glasse,
  Gordon, Greene, Güdel, Henning, Justtanont, Lagage, Meixner,
  Nørgaard-Nielsen, Ray, Ressler, van Dishoeck, \&
  Waelkens}]{rieke_mid-infrared_2015}
Rieke, G.~H., Wright, G.~S., Böker, T., {et~al.} 2015, Publications of the
  Astronomical Society of the Pacific, 127, 584, \dodoi{10.1086/682252}

\bibitem[{Rigby {et~al.}(2023)Rigby, Perrin, McElwain, Kimble, Friedman, Lallo,
  Doyon, Feinberg, Ferruit, Glasse, Rieke, Rieke, Wright, Willott, Colon,
  Milam, Neff, Stark, Valenti, Abell, Abney, Abul-Huda, Scott~Acton, Adams,
  Adler, Aguilar, Ahmed, Albert, Alberts, Aldridge, Allen, Altenburg, Álvarez
  Márquez, Alves~de Oliveira, Andersen, Anderson, Anderson, Argyriou,
  Armstrong, Arribas, Artigau, Arvai, Atkinson, Bacon, Bair, Banks, Barrientes,
  Barringer, Bartosik, Bast, Baudoz, Beatty, Bechtold, Beck, Bergeron,
  Bergkoetter, Bhatawdekar, Birkmann, Blazek, Blome, Boccaletti, Böker, Boia,
  Bonaventura, Bond, Bosley, Boucarut, Bourque, Bouwman, Bower, Bowers, Boyer,
  Bradley, Brady, Braun, Breda, Bresnahan, Bright, Britt, Bromenschenkel,
  Brooks, Brooks, Brown, Brown, Brown, Bunker, Burger, Bushouse, Cale, Cameron,
  Cameron, Canipe, Caplinger, Caputo, Cara, Carey, Carniani, Carrasquilla,
  Carruthers, Case, Catherine, Chance, Chapman, Charlot, Charlow, Chayer, Chen,
  Cherinka, Chichester, Chilton, Chonis, Clampin, Clark, Clark, Coe, Coleman,
  Comber, Comeau, Connolly, Cooper, Cooper, Coppock, Correnti, Cossou, Coulais,
  Coyle, Cracraft, Curti, Cuturic, Davis, Davis, Dean, DeLisa, deMeester,
  Dencheva, Dencheva, DePasquale, Deschenes, Hunor~Detre, Diaz, Dicken,
  DiFelice, Dillman, Dixon, Doggett, Donaldson, Douglas, DuPrie, Dupuis,
  Durning, Easmin, Eck, Edeani, Egami, Ehrenwinkler, Eisenhamer, Eisenhower,
  Elie, Elliott, Elliott, Ellis, Engesser, Espinoza, Etienne, Etxaluze, Falini,
  Feeney, Ferry, Filippazzo, Fincham, Fix, Flagey, Florian, Flynn, Fontanella,
  Ford, Forshay, Fox, Franz, Fu, Fullerton, Galkin, Galyer, García~Marín,
  Gardner, Gardner, Garland, Garrett, Gasman, Gaspar, Gaudreau, Gauthier,
  Geers, Geithner, Gennaro, Giardino, Girard, Giuliano, Glassmire, Glauser,
  Glazer, Godfrey, Golimowski, Gollnitz, Gong, Gonzaga, Gordon, Gordon,
  Goudfrooij, Greene, Greenhouse, Grimaldi, Groebner, Grundy, Guillard, Gutman,
  Ha, Haderlein, Hagedorn, Hainline, Haley, Hami, Hamilton, Hammel, Hansen,
  Harkins, Harr, Hart, Hart, Hartig, Hashimoto, Haskins, Hathaway, Havey,
  Hayden, Hecht, Heller-Boyer, Henriques, Henry, Hermann, Hernandez, Hesman,
  Hicks, Hilbert, Hines, Hoffman, Holfeltz, Holler, Hoppa, Hott, Howard,
  Howard, Hunter, Hunter, Hurst, Husemann, Hustak, Ilinca~Ignat, Illingworth,
  Irish, Jackson, Jahromi, Jakobsen, James, James, Januszewski, Jenkins,
  Jirdeh, Johnson, Johnson, Jones, Jones, Jones, Jones, Jordan, Jordan,
  Jurczyk, Jurling, Kaleida, Kalmanson, Kammerer, Kang, Kao, Karakla, Kavanagh,
  Kelly, Kendrew, Kennedy, Kenny, Keski-kuha, Keyes, Kidwell, Kinzel, Kirk,
  Kirkpatrick, Kirshenblat, Klaassen, Knapp, Scott~Knight, Knollenberg,
  Koehler, Koekemoer, Kovacs, Kulp, Kumari, Kyprianou, La~Massa, Labador,
  Labiano, Lagage, Lajoie, Lallo, Lam, Lamb, Lambros, Lampenfield, Langston,
  Larson, Law, Lawrence, Lee, Leisenring, Lepo, Leveille, Levenson, Levine,
  Levy, Lewis, Lewis, Libralato, Lightsey, Link, Liu, Lo, Lockwood, Logue,
  Long, Long, Loomis, Lopez-Caniego, Lorenzo~Alvarez, Love-Pruitt, Lucy,
  Luetzgendorf, Maghami, Maiolino, Major, Malla, Malumuth, Manjavacas,
  Mannfolk, Marrione, Marston, Martel, Maschmann, Masci, Masciarelli,
  Maszkiewicz, Mather, McKenzie, McLean, McMaster, Melbourne, Meléndez,
  Menzel, Merz, Meyett, Meza, Miskey, Misselt, Moller, Morrison, Morse,
  Moseley, Mosier, Mountain, Mueckay, Mueller, Mullally, Murphy, Murray,
  Murray, Mustelier, Muzerolle, Mycroft, Myers, Myrick, Nanavati, Nance, Nayak,
  Naylor, Nelan, Nickson, Nielson, Nieto-Santisteban, Nikolov, Noriega-Crespo,
  O’Shaughnessy, O’Sullivan, Ochs, Ogle, Oleszczuk, Olmsted, Osborne,
  Ottens, Owens, Pacifici, Pagan, Page, Park, Parrish, Patapis, Paul, Pauly,
  Pavlovsky, Pedder, Peek, Pena-Guerrero, Penanen, Perez, Perna, Perriello,
  Phillips, Pietraszkiewicz, Pinaud, Pirzkal, Pitman, Piwowar, Platais, Player,
  Plesha, Pollizi, Polster, Pontoppidan, Porterfield, Proffitt, Pueyo, Pulliam,
  Quirt, Quispe~Neira, Ramos~Alarcon, Ramsay, Rapp, Rapp, Rauscher,
  Ravindranath, Rawle, Regan, Reichard, Reis, Ressler, Rest, Reynolds, Rhue,
  Richon, Rickman, Ridgaway, Ritchie, Rix, Robberto, Robinson, Robinson,
  Robinson, Rock, Rodriguez, Rodriguez Del~Pino, Roellig, Rohrbach, Roman,
  Romelfanger, Rose, Roteliuk, Roth, Rothwell, Rowlands, Roy, Royer, Royle,
  Rui, Rumler, Runnels, Russ, Rustamkulov, Ryden, Ryer, Sabata, Sabatke, Sabbi,
  Samuelson, Sapp, Sappington, Sargent, Sauer, Scheithauer, Schlawin, Schlitz,
  Schmitz, Schneider, Schreiber, Schulze, Schwab, Scott, Sembach, Shanahan,
  Shaughnessy, Shaw, Shawger, Shay, Sheehan, Shen, Sherman, Shiao, Shih,
  Shivaei, Sienkiewicz, Sing, Sirianni, Sivaramakrishnan, Skipper, Sloan,
  Slocum, Slowinski, Smith, Smith, Smith, Smith, Snyder, Soh, Tony~Sohn, Soto,
  Spencer, Stallcup, Stansberry, Starr, Starr, Stewart, Stiavelli, Straughn,
  Strickland, Stys, Summers, Sun, Sunnquist, Swade, Swam, Swaters, Swoish,
  Taylor, Taylor, Te~Plate, Tea, Teague, Telfer, Temim, Thatte, Thompson,
  Thompson, Thomson, Tikkanen, Tippet, Todd, Toolan, Tran, Trejo, Truong,
  Tsukamoto, Tustain, Tyra, Ubeda, Underwood, Uzzo, Van~Campen, Vandal,
  Vandenbussche, Vila, Volk, Wahlgren, Waldman, Walker, Wander, Warfield,
  Warner, Wasiak, Watkins, Weaver, Weilert, Weiser, Weiss, Weissman, Welty,
  West, Wheate, Wheatley, Wheeler, White, Whiteaker, Whitehouse, Whiteleather,
  Whitman, Williams, Willmer, Willoughby, Wilson, Wirth, Wislowski, Wolf,
  Wolfe, Wolff, Workman, Wright, Wu, Wu, Wymer, Yates, Yeager, Yeates, Yerger,
  Yoon, Young, Yu, Zak, Zeidler, Zhou, Zielinski, Zincke, \&
  Zonak}]{rigby_science_2023}
Rigby, J., Perrin, M., McElwain, M., {et~al.} 2023, Publications of the
  Astronomical Society of the Pacific, 135, 048001,
  \dodoi{10.1088/1538-3873/acb293}

\bibitem[{Roellig {et~al.}(2004)Roellig, Cleve, Sloan, Wilson, Saumon, Leggett,
  Marley, Cushing, Kirkpatrick, Mainzer, \& Houck}]{roellig_spitzer_2004}
Roellig, T.~L., Cleve, J. E.~V., Sloan, G.~C., {et~al.} 2004, The Astrophysical
  Journal Supplement Series, 154, 418, \dodoi{10.1086/421978}

\bibitem[{Saumon {et~al.}(2000)Saumon, Geballe, Leggett, Marley, Freedman,
  Lodders, B.~Fegley, \& Sengupta}]{saumon_molecular_2000}
Saumon, D., Geballe, T.~R., Leggett, S.~K., {et~al.} 2000, The Astrophysical
  Journal, 541, 374, \dodoi{10.1086/309410}

\bibitem[{Schneider {et~al.}(2015)Schneider, Cushing, Kirkpatrick, Gelino,
  Mace, Wright, Eisenhardt, Skrutskie, Griffith, \&
  Marsh}]{schneider_hubble_2015}
Schneider, A.~C., Cushing, M.~C., Kirkpatrick, J.~D., {et~al.} 2015, The
  Astrophysical Journal, 804, 92, \dodoi{10.1088/0004-637X/804/2/92}

\bibitem[{Skemer {et~al.}(2016)Skemer, Morley, Allers, Geballe, Marley,
  Fortney, Faherty, Bjoraker, \& Lupu}]{skemer_first_2016}
Skemer, A.~J., Morley, C.~V., Allers, K.~N., {et~al.} 2016, The Astrophysical
  Journal Letters, 826, L17, \dodoi{10.3847/2041-8205/826/2/L17}

\bibitem[{Visscher {et~al.}(2006)Visscher, Lodders, \&
  Bruce~Fegley}]{visscher_atmospheric_2006}
Visscher, C., Lodders, K., \& Bruce~Fegley, J. 2006, The Astrophysical Journal,
  648, 1181, \dodoi{10.1086/506245}

\bibitem[{Werner {et~al.}(2004)Werner, Roellig, Low, Rieke, Rieke, Hoffmann,
  Young, Houck, Brandl, Fazio, Hora, Gehrz, Helou, Soifer, Stauffer, Keene,
  Eisenhardt, Gallagher, Gautier, Irace, Lawrence, Simmons, Cleve, Jura,
  Wright, \& Cruikshank}]{werner_spitzer_2004}
Werner, M.~W., Roellig, T.~L., Low, F.~J., {et~al.} 2004, The Astrophysical
  Journal Supplement Series, 154, 1, \dodoi{10.1086/422992}

\bibitem[{Wright {et~al.}(2010)Wright, Eisenhardt, Mainzer, Ressler, Cutri,
  Jarrett, Kirkpatrick, Padgett, McMillan, Skrutskie, Stanford, Cohen, Walker,
  Mather, Leisawitz, Gautier, McLean, Benford, Lonsdale, Blain, Mendez, Irace,
  Duval, Liu, Royer, Heinrichsen, Howard, Shannon, Kendall, Walsh, Larsen,
  Cardon, Schick, Schwalm, Abid, Fabinsky, Naes, \&
  Tsai}]{wright_wide-field_2010}
Wright, E.~L., Eisenhardt, P. R.~M., Mainzer, A.~K., {et~al.} 2010, The
  Astronomical Journal, 140, 1868, \dodoi{10.1088/0004-6256/140/6/1868}

\bibitem[{Yamamura {et~al.}(2010)Yamamura, Tsuji, \&
  Tanabé}]{yamamura_akari_2010}
Yamamura, I., Tsuji, T., \& Tanabé, T. 2010, The Astrophysical Journal, 722,
  682, \dodoi{10.1088/0004-637X/722/1/682}

\end{thebibliography}
\bibliographystyle{aasjournal}

\end{document}